\documentclass[namedreferences]{SolarPhysics}
\usepackage[optionalrh,solaenum]{spr-sola-addons} 
\usepackage{graphicx}        
\usepackage{color}           
\usepackage{url}             
\usepackage{epstopdf}

\newcommand{\aap}{    {\it Astron. Astrophys.}}

\newcommand{\apj}{    {\it Astrophys. J.}}
\newcommand{\apjl}{   {\it Astrophys. J. Lett.}}

\newcommand{\jgr}{    {\it J. Geophys. Res.}}

\newcommand{\nat}{    {\it Nature}}

\newcommand{\solphys}{{\it Solar Phys.}}

\newcommand{\ssr}{    {\it Space Sci. Rev.}}

\newcommand\ion[2]{#1$\;${\scshape{#2}}}

\newcommand{\HeII}{\ion{He}{ii}}
\newcommand{\Halpha}{H$\alpha$}

\newcommand{\FeIX}{\ion{Fe}{ix}}
\newcommand{\FeXII}{\ion{Fe}{xii}}

\newcommand{\kms}{km~s$^{-1}$}
\newcommand{\degree}{\ensuremath{^\circ}}

\def\arcsec{\hbox{$^{\prime\prime}$}}

\begin{document}

\begin{article}

\begin{opening}

\title{On the structure and evolution of a polar crown prominence/filament system\\ {\it Solar Physics}}

\author{N.~\surname{K.~Panesar}$^{1}$,$^{2}$\sep
        D.~\surname{E.~Innes}$^{1}$\sep
        D.~\surname{J.~Schmit}$^{1}$\sep
        S.~\surname{K.~Tiwari}$^{1}$,$^{3}$\sep
       }
\runningauthor{Panesar \textit{et al.}}
\runningtitle{Observations of a polar crown prominence}

   \institute{$^{1}$ {Max-Planck Institut f\"{u}r Sonnensystemforschung, Max-Planck-Str. 2, 37191
   Katlenburg-Lindau}\\
                     email:
                     \url{panesar@mps.mpg.de} \\
              $^{2}$ {Institut f\"{u}r Astrophysik G\"{o}ttingen, Physik Fakult\"{a}t, Friedrich-Hund-Platz 1, D-37077
              G\"{o}ttingen}\\
              $^{3}${NASA Marshall Space Flight Center, ZP 13, Huntsville, AL 35812, USA } \\
                    }

\begin{abstract}
   Polar crown prominences are made of chromospheric plasma partially circling the Sun's poles between 60\degree\ and 70\degree\
   latitude. We aim to diagnose the 3D dynamics of a polar crown prominence
   using high cadence EUV images from the \textit{Solar Dynamics Observatory} (SDO)/AIA at
   304 and 171 \AA\ and the \textit{Ahead} spacecraft of the \textit{Solar Terrestrial Relations
   Observatory} (STEREO-A)/EUVI at 195 \AA. Using time series across specific
   structures we compare flows across the disk in 195 \AA\ with the
   prominence dynamics seen on the limb. The
   densest prominence material forms vertical columns which are
   separated by many tens of Mm and connected by dynamic bridges of plasma
 that are clearly visible in 304/171 \AA\ two-color images.
We also observe intermittent but repetitious
   flows with velocity 15 km s$^{-1}$ in the prominence that appear to be associated with EUV
   bright points on the solar disk. The boundary between the prominence and the overlying cavity appears as a sharp
   edge. We discuss the structure of the coronal cavity seen both above and
around the prominence. SDO/HMI and GONG magnetograms are used
   to infer the underlying magnetic topology. The evolution and structure of the prominence with
   respect to the magnetic field seems to agree with the filament linkage model.
\end{abstract}
\end{opening}


\section{Introduction}

\begin{figure*}
   \centering
 \includegraphics[width=\linewidth]{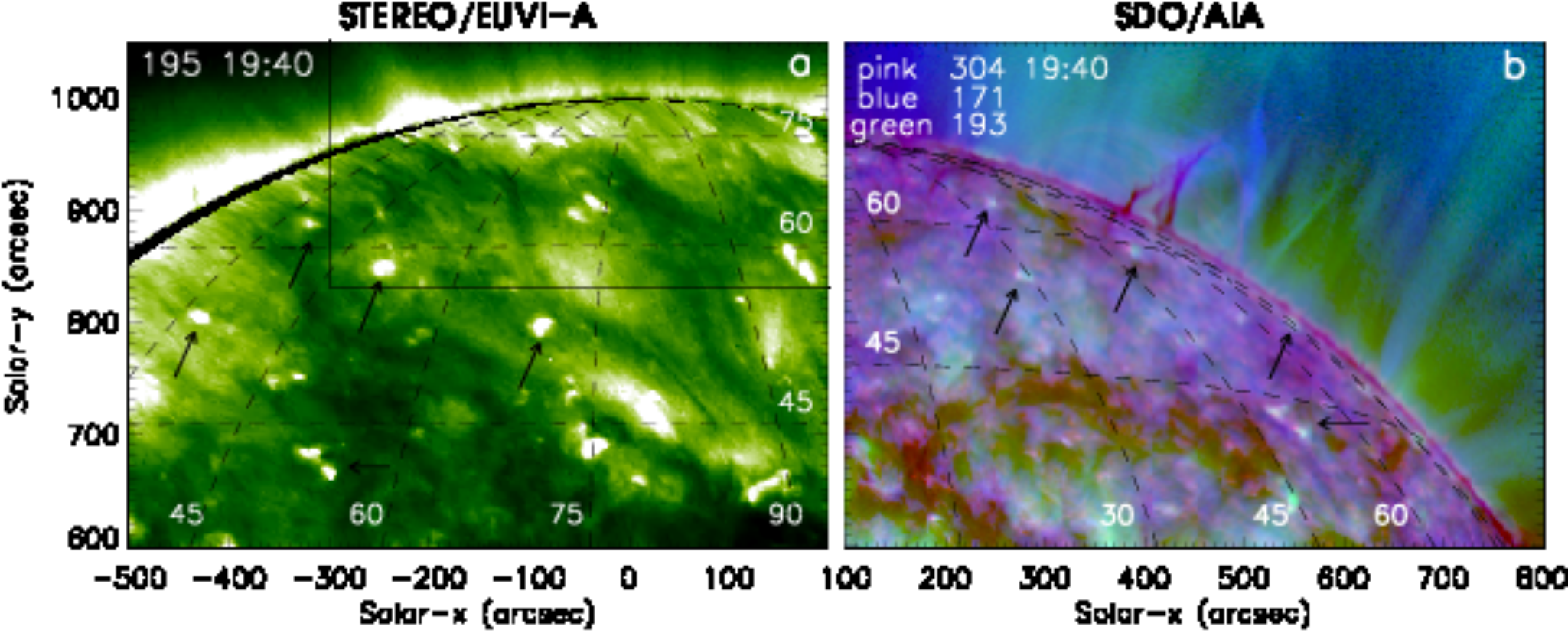}
     \caption{a) Prominence/filament system observed on 13 February 2011 (a) STEREO-A 195 \AA\ b) SDO (304,
      171, 193 \AA\ composite). Arrows
  point to common bright points in the STEREO and SDO images. Grid lines are SDO longitude and latitude separated by 15\degree.
  The righthand longitude line shows the position of SDO limb. The lowest latitude line is 45\degree.
  The STEREO image has been derotated to align with disk features at 00:05 UT on 15 February.
  A black box outlines the region presented in Figure~\ref{footpoints}.}
\label{over}
\end{figure*}

As their name suggests, polar crown prominence/filament systems create a crown that approximately outlines the polar coronal holes
\cite{hir85,tand95}. They form between dispersed active region flux and
polar coronal holes. Their position and number changes with the solar cycle \cite{Webb84,McIntosh92}.
Starting a few years before solar maximum, the phase in which the reported observations were made,
polar crown filaments become more common \cite{zir97,Minar98}.
On the disk they appear as rows of dark pillars or vertical columns with intermittent filamentary connections \cite{lin2003,sch10,dud12,li2013}.
On the limb, persistent small-scale flows are seen in the cold plasma pillars \cite{cha08,Berger08,Berger11}
that often reside in long-lived coronal cavities \cite{gib06,gib10,hab10,don11}.

One key to understanding the formation of polar crown prominences is the linkage of flux
across and between dispersing active regions \cite{Martens01,mac01}.
Most filaments are found between initially unrelated active
regions \cite{gai97,Martin98,Mackay08} along the magnetic polarity
inversion lines (PILs) which separate the regions of opposite polarity magnetic fields \cite{Martin73}. It is probable
that magnetic reconnection, driven by flows converging to the PIL, plays an important role in the
filament formation and maintaining the filament mass (\textit{e.g.} \opencite{cha03}).
 As new connections are made between the
adjacent active regions, open flux regions develop along the outer edges of
the original bipoles \cite{mac06}.

Prominences and their on-disk counterparts, filaments,
have been thoroughly investigated with imaging and spectroscopic
techniques (\inlinecite{lab10} and references therein).
 They are relatively cool and dense structures suspended in the
hot million-degree kelvin corona \cite{hir85,Mackay10,lab10}. When observed in chromospheric lines,
they appear to consist of two main structures: the spine that runs along the filament channel,
and the barbs that project sideways from the spine and connect to the photosphere
\cite{Martin98,aul98}.
Often in images of polar crown filaments only the barbs, sometimes known as the filament feet, are visible as dark pillars
\cite{sch10,dud12,li2013}.
 The two main reasons for prominences appearing dark in EUV images is
absorption and/or volume blocking \cite{anz05}.
Absorption is due to the photoionization of H and He, by the coronal
radiation passing through the prominence \cite{kuc98,hei01,hei03}.
Volume blocking is caused by the absence of coronal plasma in a
volume of space which does not participate in the emission of coronal lines
\cite{sch04,hei08}. Some of the prominence features can also be visible as emission in the 171 \AA\ filter
\cite{par12}.

Prominences observed in \Halpha\ and \HeII\ have different morphological
features \cite{wangh98}.
In observations, the \Halpha\ emission coincides with the 193\AA\ absorption because
the optical thickness of the prominence observed in the \FeXII\ line is similar
to the optical thickness in the \Halpha\ \cite{anz05}.
 Spectroscopic observations of lines
formed at transition region and coronal temperatures show plasma streaming
through the corona with velocities up to 70 km s$^{-1}$ for typically 13
min \cite{kuc03}. Such flows would therefore travel a distance of 50 000 km, equivalent to a
supergranule cell diameter. One of the outstanding problems is to disentangle
structures along the line-of-sight and to see the connections between the
cool \Halpha~and 304 \AA\ emission and the hot coronal emission.

An important and probably necessary condition for the stability of the prominence/filament system
is an overlying coronal arcade above the cold prominence
plasma and related coronal cavity \cite{low95,Hudson99,Gibson06,hei08}.
The arcade is nearly potential field and surrounds both the cavity
and the prominence. Most models for the cavity and prominence invoke either a flux-rope
\cite{kup74,Priest89,Balle89,low95,aul98,low02} or a sheared-arcade \cite{ant94,dev00,aul02}. In these models the
dense prominence plasma lies in the dips of the field and the cavity is filled with the surrounding, low-density plasma.
 Typically the transverse fields of quiescent polar crown prominences are observed to have an
inverse polarity with respect to their overlying arcade
\cite{leroy84,anz94,ant94,kup96}. This is consistent with both the flux-rope and
sheared arcade models. Coronal magnetic field measurements favour a
flux-rope configuration in at least one of the observed polar crown cavities
\cite{rach13,bak13}.

The cavity in extreme ultraviolet (EUV) images is sometimes seen
 on top of cooler prominence material with horns of cooler
material outlining the base of the cavity \cite{Regnier11,schmit13,panesar13}, and sometimes surrounding the prominence
\cite{berger12}. From the top, the cavity shows up as a dim region in EUV images \cite{vas09}.

In the present work, we focus on the dynamics of a small polar crown prominence filament  system seen on the north-west limb
from Earth using high cadence \textit{Solar Dynamics Observatory}
(SDO) images from the \textit{Atmospheric Imaging Assembly} (AIA) \cite{lem12} and near the central meridian in 195 \AA\
\textit{Solar Terrestrial Relations Observatory Ahead} (STEREO-A) images from
the \textit{Extreme UltraViolet Imager} (EUVI) \cite{Howard08}.
The prominence is typical of the many small ones seen at high latitudes.
Hazy 171 \AA\ emission extends from the top and sides of the main 304 \AA\ prominence pillars, into and from which plasma streams along spidery legs.
Recently, \inlinecite{su12} have also analysed SDO and STEREO quadrature observations to investigate the magnetic structure
of a large polar crown prominence/filament system, in the days preceding its partial eruption.
Here we use the stereoscopic perspective of SDO and STEREO-A to separate the filament into pillars and interconnecting flows.
In particular we are able to use combined 304 and 171 \AA\ AIA images to track the flows through the corona,
and to find the starting and end points of the flows on the disk.
Using the two instruments in quadrature, we are also able to investigate the events leading to the
appearance of EUV cavities around and over the prominence.
To identify the magnetic configurations underlying the prominence, we have used line-of-sight magnetograms
obtained by the \textit{Helioseismic and Magnetic Imager} (HMI) \cite{sch12} and the \textit{Global Oscillation Network Group} (GONG) \cite{harvey96}.

The plan of the paper is as follows: In Section 2, we  describe the observations, and  give details of the data
analysis; In Section 3, we first give an overview of the polar crown prominence and the structure of the filament channel.
Then we investigate the evolution of the prominence over about 40 hours so as to identify the main prominence pillars in the
STEREO-A images. Finally in Section 4 we present a sketch of the prominence/filament system and relate
it to the large-scale magnetic field configuration.

\begin{figure*}[ht]
   \centering
   \includegraphics[width=0.9\linewidth]{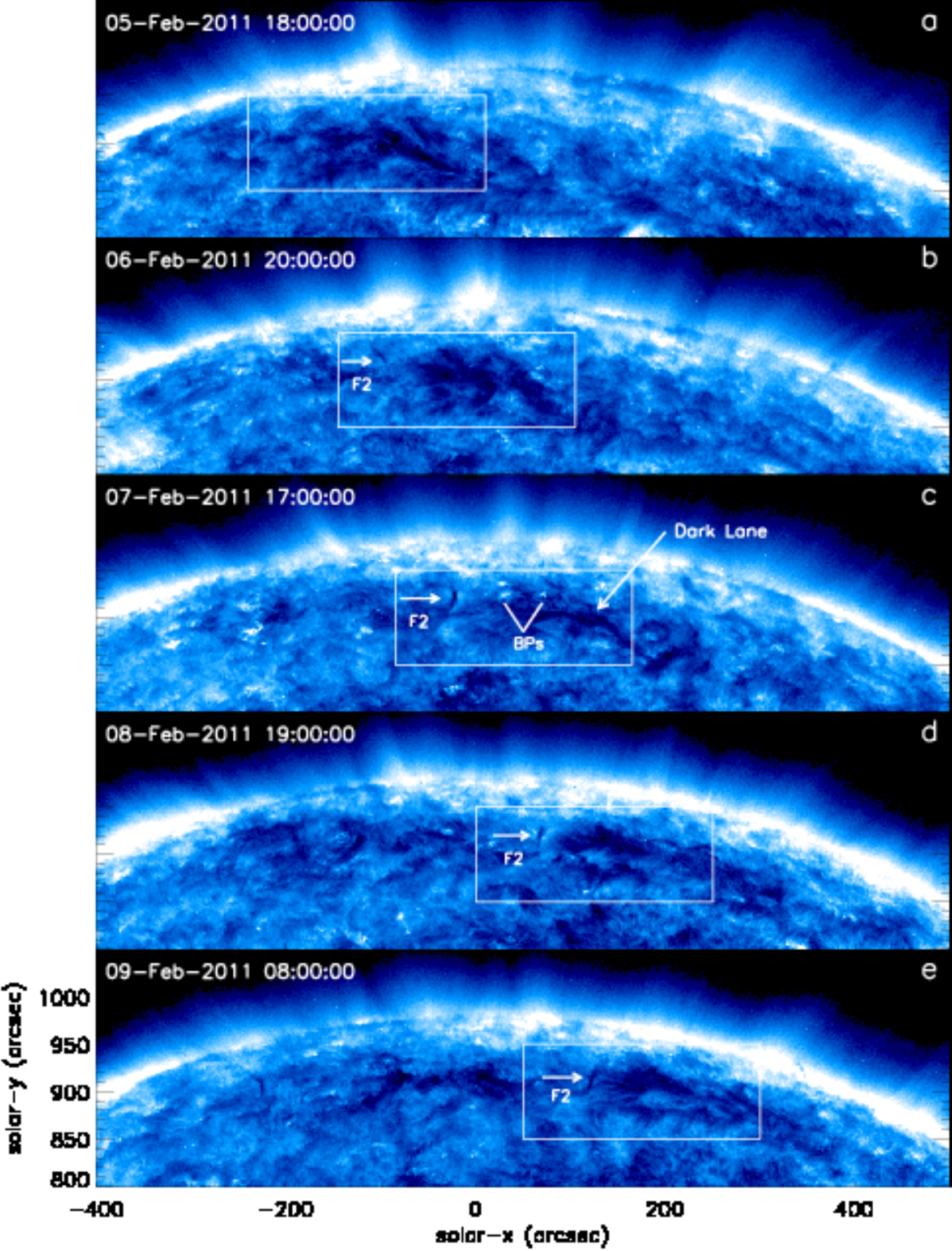}
   \caption{Evolution of the filament system from 5--9 February 2011:
   Sequences of SDO/AIA 171 \AA\ intensity images showing the growth of pillar F2 and  bright points (BPs) in the filament channel.
   The white box roughly outlines the region of the filament channel observed on the limb on 13 and 14 February 2011.
   The dark lane is indicated with an arrow in (c).}
\label{evolution}
\end{figure*}

\begin{figure*}
   \centering
   \includegraphics[width=0.9\linewidth]{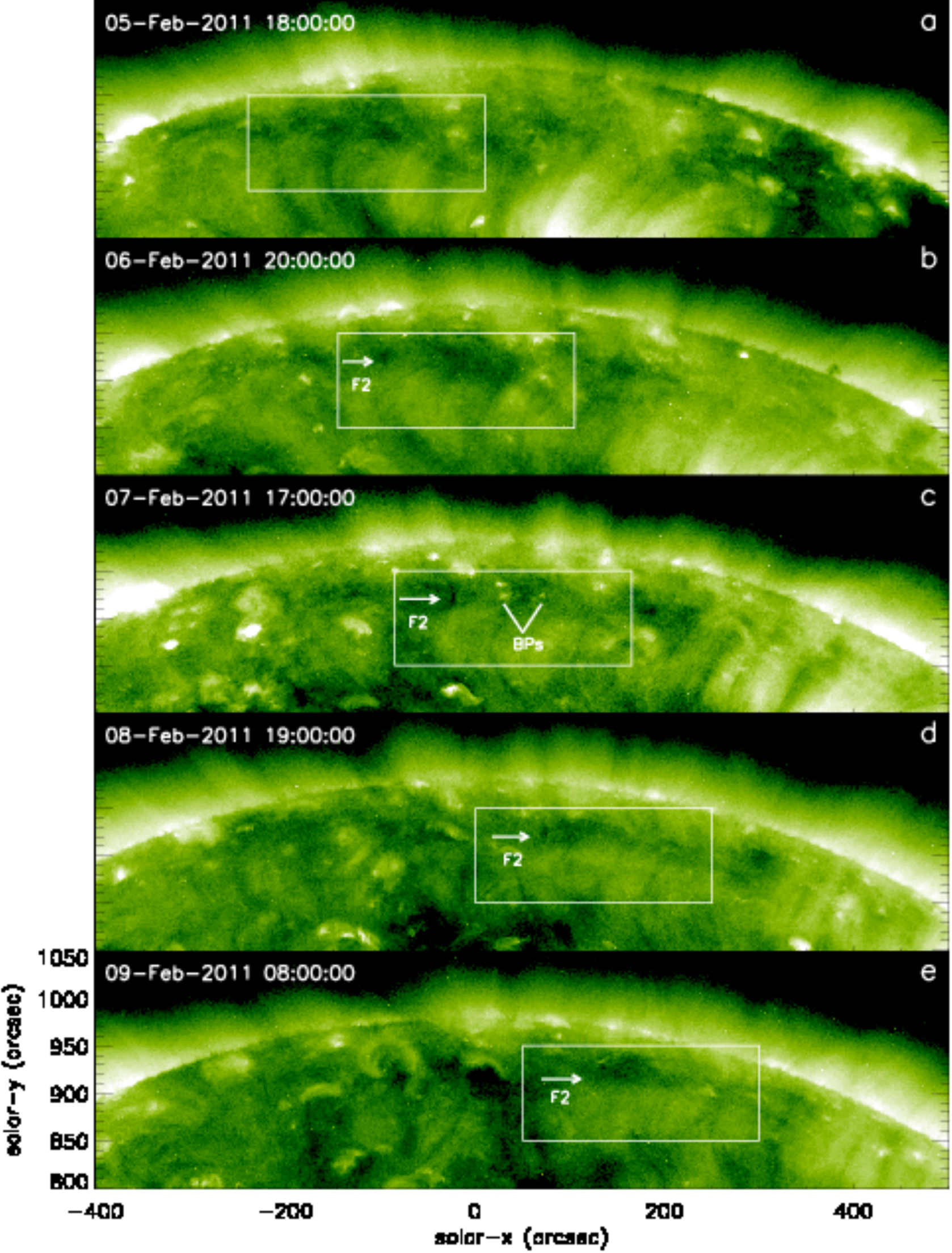}
   \caption{Evolution of the filament system from 5--9 February 2011:
   Sequences of SDO/AIA 193 \AA\ intensity images showing the pillar F2 and bright points (BPs) in the filament channel.
   The images are taken at the
   same time as those in Figure~\ref{evolution}.}
\label{evo}
\end{figure*}

\begin{figure*}[ht]
   \centering
   \includegraphics[width=0.9\linewidth]{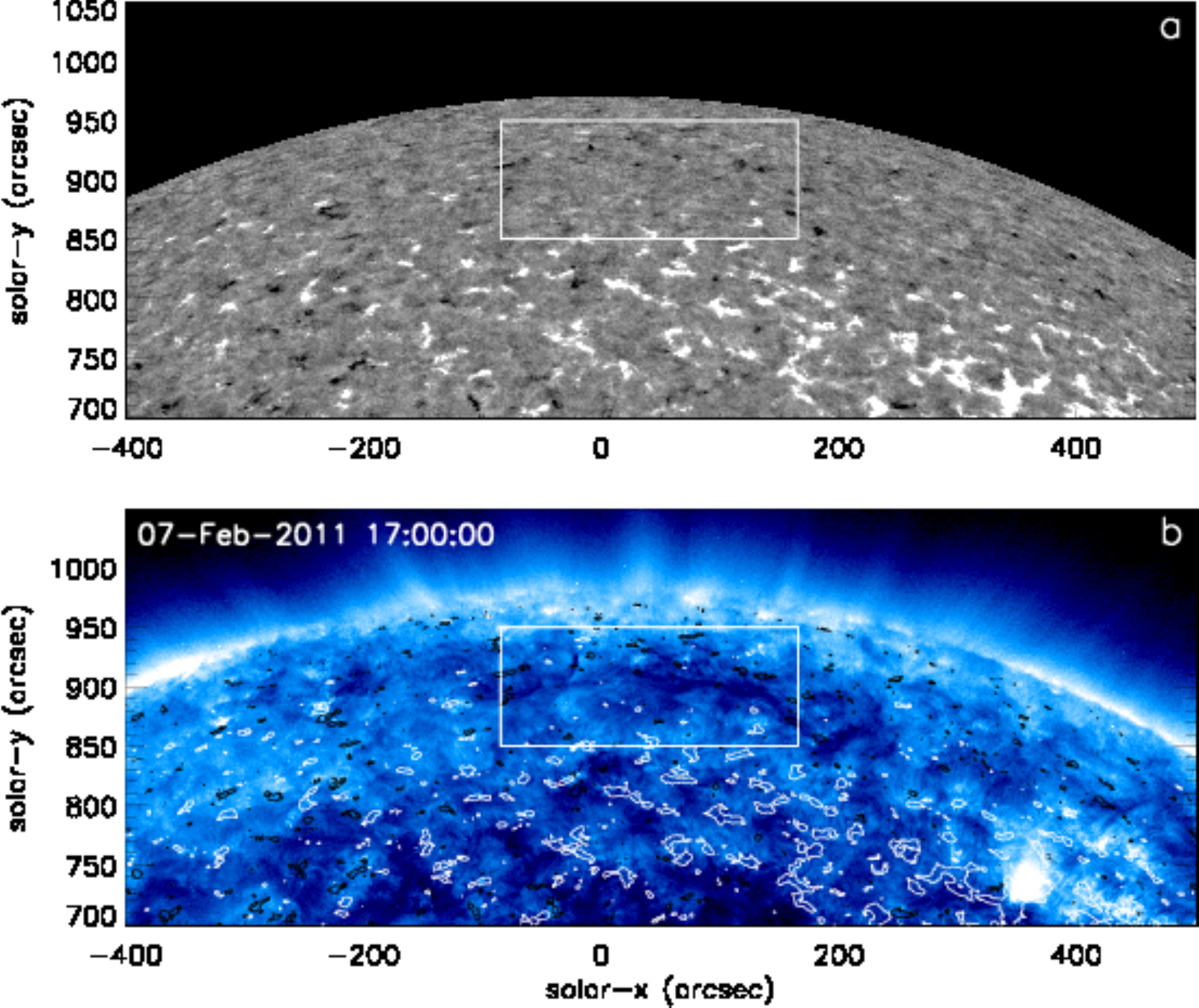}
   \caption{Magnetic field structure on 7 February 2011: (a) SDO/HMI 6173 \AA\ averaged magnetogram;
   (b) SDO/AIA 171 \AA\ intensity image overlaid with contours of
   $\pm $~10 G.}
\label{hmi}
\end{figure*}

\section{Observations and data analysis} 

The prominence studied here was seen from Earth (SDO) on the north-west limb of the Sun between latitudes 60-70\degree\ on 13 and 14 February 2011.
During our observations, the separation angle between SDO and STEREO-A was 86\degree\,
so that the prominence was close to the central meridian in STEREO-A images.

AIA provides high spatial resolution
(0.6 arcsec pixel$^{-1}$) full disk images with a cadence of 12~s in 10 wavelength bands,
including seven extreme ultraviolet (EUV), two ultraviolet and one
visible filter \cite{lem12}.
For the SDO analysis, we select three channels: 304 \AA\ which is dominated by the \HeII\ lines formed
 between $5-8\times10^{4}$~K, 171 \AA\ which is centered on the \FeIX\ line formed around
 $6.5-8\times10^{5}$~K, and 193 \AA\ which is centered on the \FeXII\ line
 formed between $1-2\times10^{6}$~K  \cite{del11,lem12,par12}.
 The prominence flows are best seen in 304 \AA\ and the overlying hotter
plasma and cavity in 171 \AA\ and 193 \AA.

To enhance the visibility of the 171 \AA\ and 193 \AA\ structures, we removed
an average coronal background from the images. The background is computed as the
median of two months data - January and February, 2011. This gave a smooth background
with intensity decreasing radially outwards from the limb, as well as
the correct average angular dependence. The displayed images are the logarithm
of the ratio of the images to the background: both taken at the same solar coordinates.
We found that this gave better contrast
than could be obtained by straight subtraction of the background image, because
it reflects the relative rather than absolute changes. Two-colour images are
constructed by combining scaled 304 \AA\ intensity and 171 \AA\ background ratioed
images. Because the standard AIA colours for 193, 171, and 304 \AA\ filters are not easy to distinguish in 3- or 2-color
images, we have used the green color for 195 and 193 \AA, blue for the 171 \AA\  and red for the 304 \AA\ images.

We have used STEREO-A 195 \AA\ images with a time cadence of 5 minutes.
For most of the analysis the images have been derotated to a particular
time and SDO grid lines are superimposed so that the position of the SDO limb can be easily identified.
 We set the solar B angle to zero only for the on-disk features (STEREO images)
because this makes the alignment with the off-limb features easier. Setting
 the solar B angle to zero in the off-limb images does not change the alignment of the prominence
features because they are superimposed. To relate the 195 \AA\ images to structures in the filament channel we looked at SDO 304, 171 and 193 \AA\
images taken one week earlier, 5--9 February,  when the prominence appeared as a filament near the central meridian from SDO.

We also used SDO/HMI magnetograms taken on the 7 February to look at the magnetic field under the filament channel.
HMI provides full disk, line-of-sight magnetograms with a resolution of 0.5 arcsec pixel$^{-1}$ and a cadence of
45~s \cite{sch12}. Since the field was very weak, we built up a deep magnetogram for the  7~February
 2011 by averaging 48 coaligned magnetograms taken throughout the day.  The large-scale magnetic structure was obtained from
GONG 360\degree\ synoptic magnetograms of the Sun, coaligned to  360\degree\ STEREO-SDO composites of the
195/193 \AA\ EUV emission on the 17 January and the 13 February.

\begin{figure*}
   \centering
   \includegraphics[width=\linewidth]{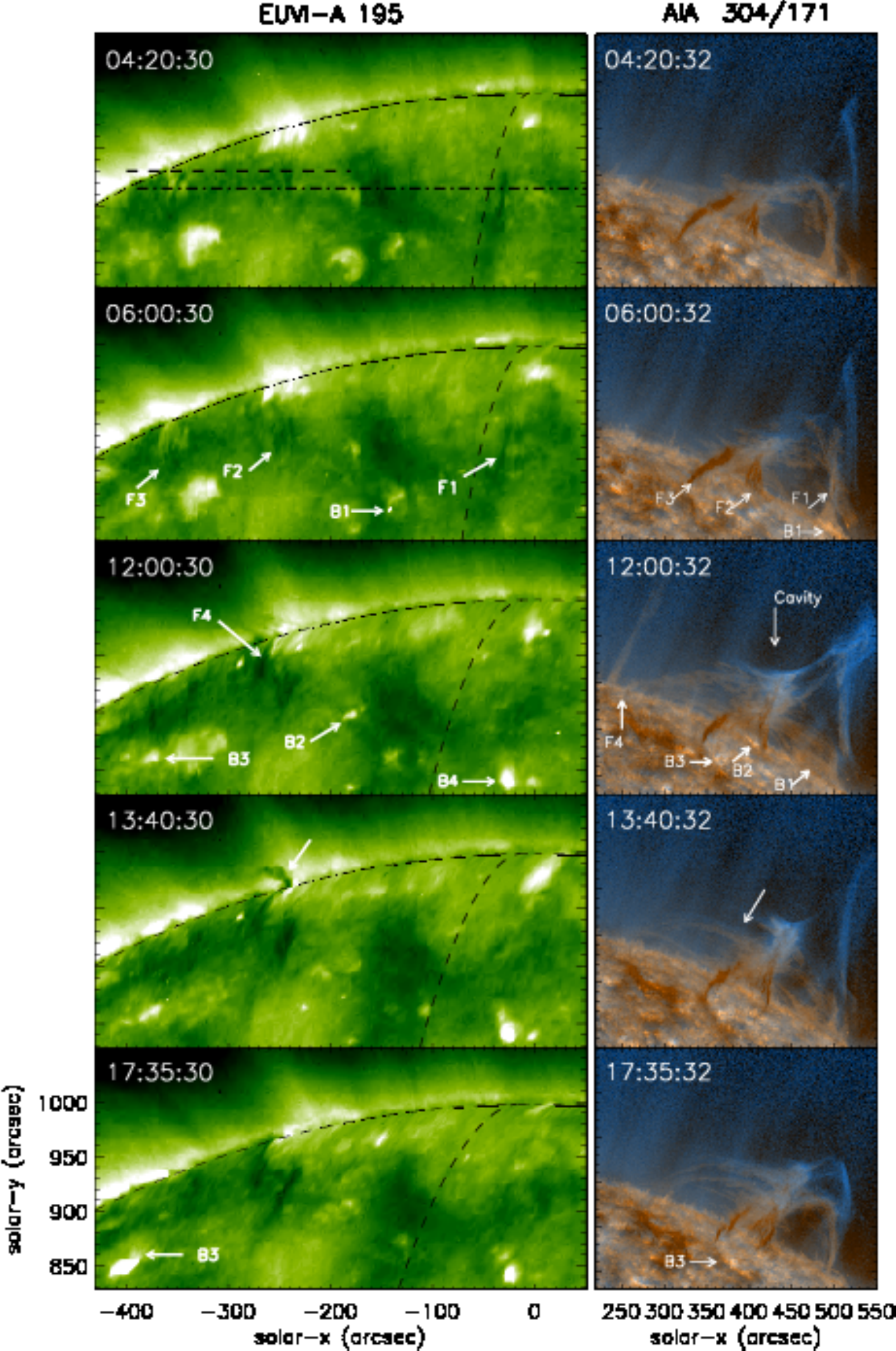}
   \caption{Time evolution of the prominence/filament system on 13 February: (left) STEREO-A 195 \AA\ intensity images derotated to 13 February 2011 00:05 UT; (right) corresponding two-color SDO 304 \AA/171 \AA\ images.
   The SDO limb is marked with a dashed line on the STEREO images. F1, F2, F3 and F4 indicate the
      positions of prominence pillars. Common bright points (B1, B2, B3 and B4) are also indicated in the second and third rows, and
      the top of the spider prominence in the fourth.
      In the top left panel, the horizontal dot-dashed line shows the position of a cut through the pillars (Figure~\ref{pillarcut})
      and the horizontal dashed line the cut along the
      spider feature (Figure~\ref{spi}).
      The cavity is indicated on the SDO limb image at 12:00 UT. The Y-axis is the same in both SDO and STEREO images. These are selected frames from the movie MOVIE13.}
\label{footpoints}
\end{figure*}

\section{Overview} 

    The small polar crown prominence/filament system, as seen from the two viewing angles at 19:40~UT on 13 February 2011, is shown in Figure~\ref{over}.
    Grid lines are the same in both images and the SDO limb is indicated by the right hand dashed line on the STEREO image.
    The SDO overview image shown here includes the 193 \AA\ as well as the 304 \AA\ and 171 \AA\ in the composite because the bright points,
    used as markers, are much more visible when the 193 \AA\ emission is included.
    Several of the common bright points are indicated with arrows. The bright points east of the 60\degree\ grid
    lines are clearly identifiable in both STEREO and SDO, but those closer to the limb are partially obscured in the
    SDO image due to the column of intervening gas along the line-of-sight. The prominence which lies between
    60--75\degree\ latitude
    is difficult to see in the disk image.
 We have outlined the filament channel region with a black box. Several dark structures in the filament channel are just discernable.
 The cold plasma should be dark in the 195 \AA\
    disk image due to absorption/blocking of EUV emission by the cold, gas \cite{anz05}. However the darkening due to prominence
    material can easily be confused with darkening caused by a coronal cavity or regions of
    low coronal density in the filament
    channel. We therefore looked at  the filament channel in the SDO images from 5--9 February 2011
    to distinguish the various structures.

\subsection{Structure of the filament channel}

\begin{figure*}
   \centering
   \includegraphics[width=\linewidth]{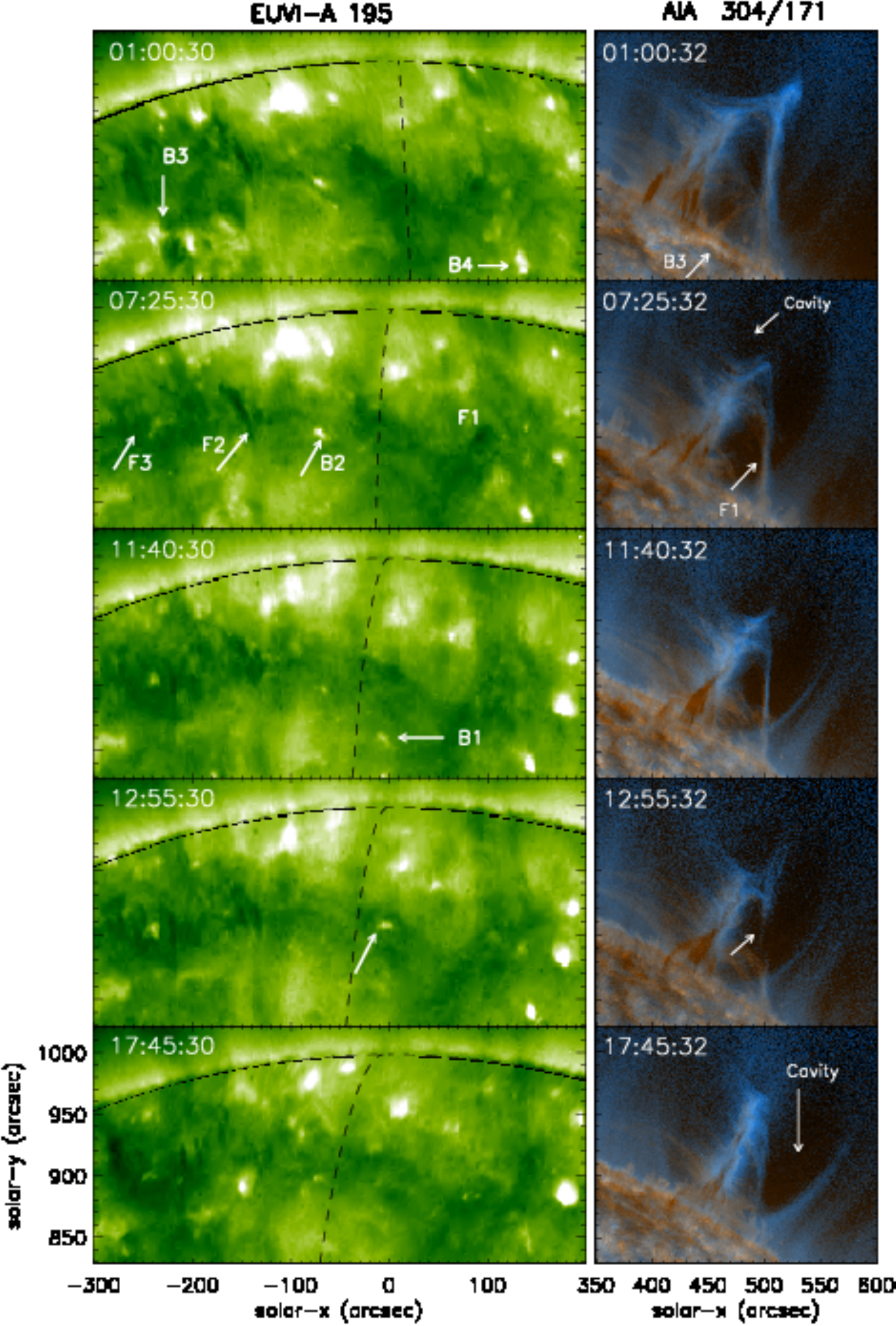}
   \caption{Evolution of the prominence/filament system on 14 February: (left) STEREO-A 195 \AA\ intensity images derotated to
   12:05 UT on 14 February 2011; (right)
   corresponding two-color SDO 304 \AA/171 \AA\ images. Annotated features are same as those in Figure~\ref{footpoints}.
   At B3 there is an eruption seen in both SDO and STEREO images that leads to a coronal dimming.
The southern pillar breaks at 12:40 UT. An arrow points at a brightening on the disk at 12:55 UT. These are selected frames from the movie MOVIE14.}
\label{cavity}
\end{figure*}

 The structure of the filament channel when it was on the disk is illustrated in Figures~\ref{evolution} and ~\ref{evo},
which show one SDO 171 \AA\ and one 193 \AA\ image per day from 5--9 February.
By comparing the two, we could identify features of the filament channel
that were clearly visible in the 171 \AA\ and 193 \AA\ images, and thus better
interpret the 195 \AA\ STEREO images.
The filament channel is seen most clearly in the 171 \AA\ images. The white box highlights the part of the filament channel that we have
analysed, when it appeared on the limb. The filament channel shows a slowly evolving
broad, dark lane in 171 \AA\ images (Figure~\ref{evolution}), which is partially
obscured in the 193 \AA\ images. On the 6 February, a pillar (marked F2) appears which, by tracking its motion across the disk,
we have identified as one of the  pillars that was seen on the SDO limb and the STEREO disk
on 13 and 14 February (Figures~\ref{footpoints} and ~\ref{cavity}).
 We also observed a region with
 small-scale bright points that produced multiple short-lived brightenings (see Figure~\ref{evolution}c) just north of the main filament
plasma. These bright points were seen simultaneously in the 193 \AA\
images (Figure~\ref{evo}c). A region of reduced emission is conspicuous in the
193 \AA\ image between three bright points in
Figure~\ref{evo}c.
By looking at the five-day sequence of 193 \AA\ images, we saw that this region was
dark throughout the 7 and 8 February. When the filament channel
 was on the SDO limb, a pronounced region of reduced 195 \AA\ emission was also noticeable in the
 STEREO images at roughly the same position with respect to the pillar.
 As discussed later, it appears that this is associated with the coronal cavity.

Figure~\ref{hmi}a represents the SDO/HMI map
of the line-of-sight magnetic field on the 7 February 2011. A white box in  Figure~\ref{hmi}a,b outlines the
region of the filament channel investigated. On the whole, the  filament channel
 coincides with a region of very low magnetic flux. Diffused active region
 positive flux is distributed to the south of the filament channel and smaller concentrations
 of negative flux are found on the north and west of the channel.

\begin{figure}
   \centering
   \includegraphics[width=\linewidth]{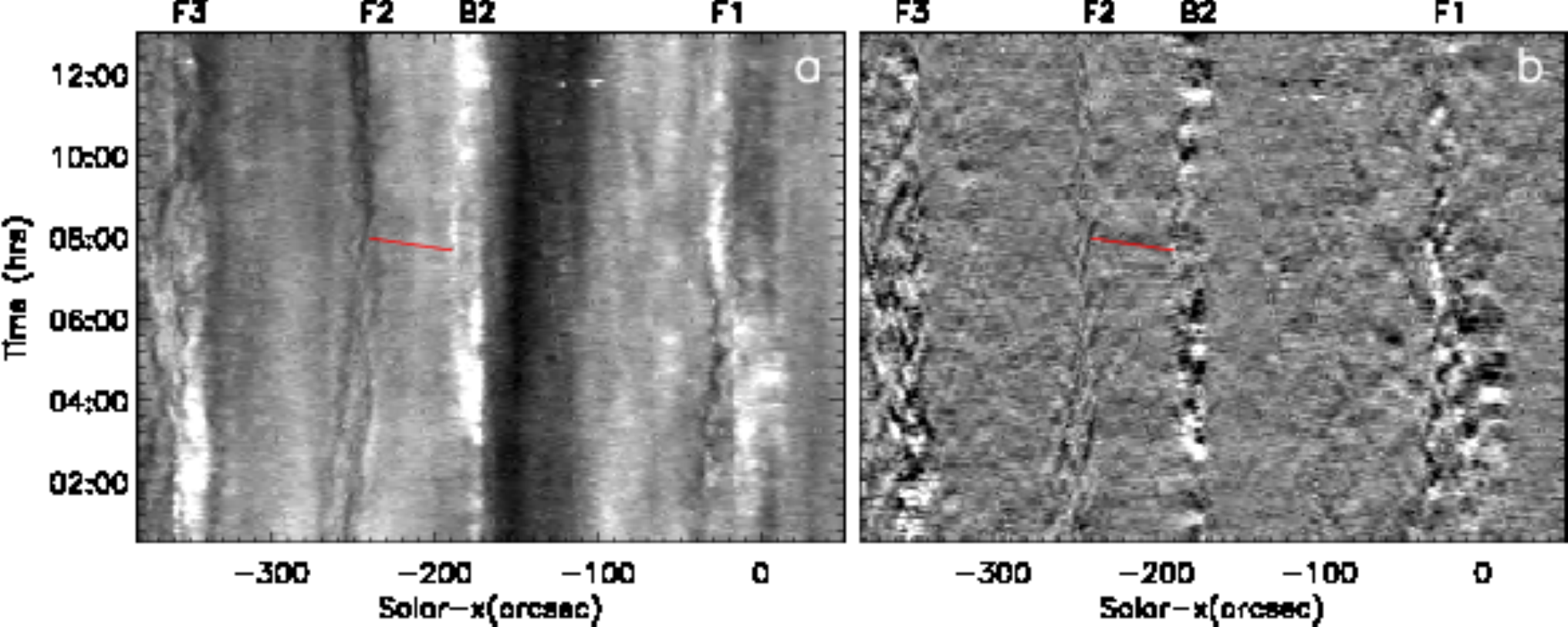}
      \caption{
        a) STEREO-A 195~\AA\ intensity time series and b) running difference time series along the dot-dashed line in Figure~\ref{footpoints} on 13 February 2011.
   F1, F2 and F3 are the positions of pillars. The red diagonal line indicates the feature between B2 and F2 where we calculate the velocity.
   }
   \label{pillarcut}
\end{figure}

\begin{figure}
   \centering
   \includegraphics[width=0.8\linewidth]{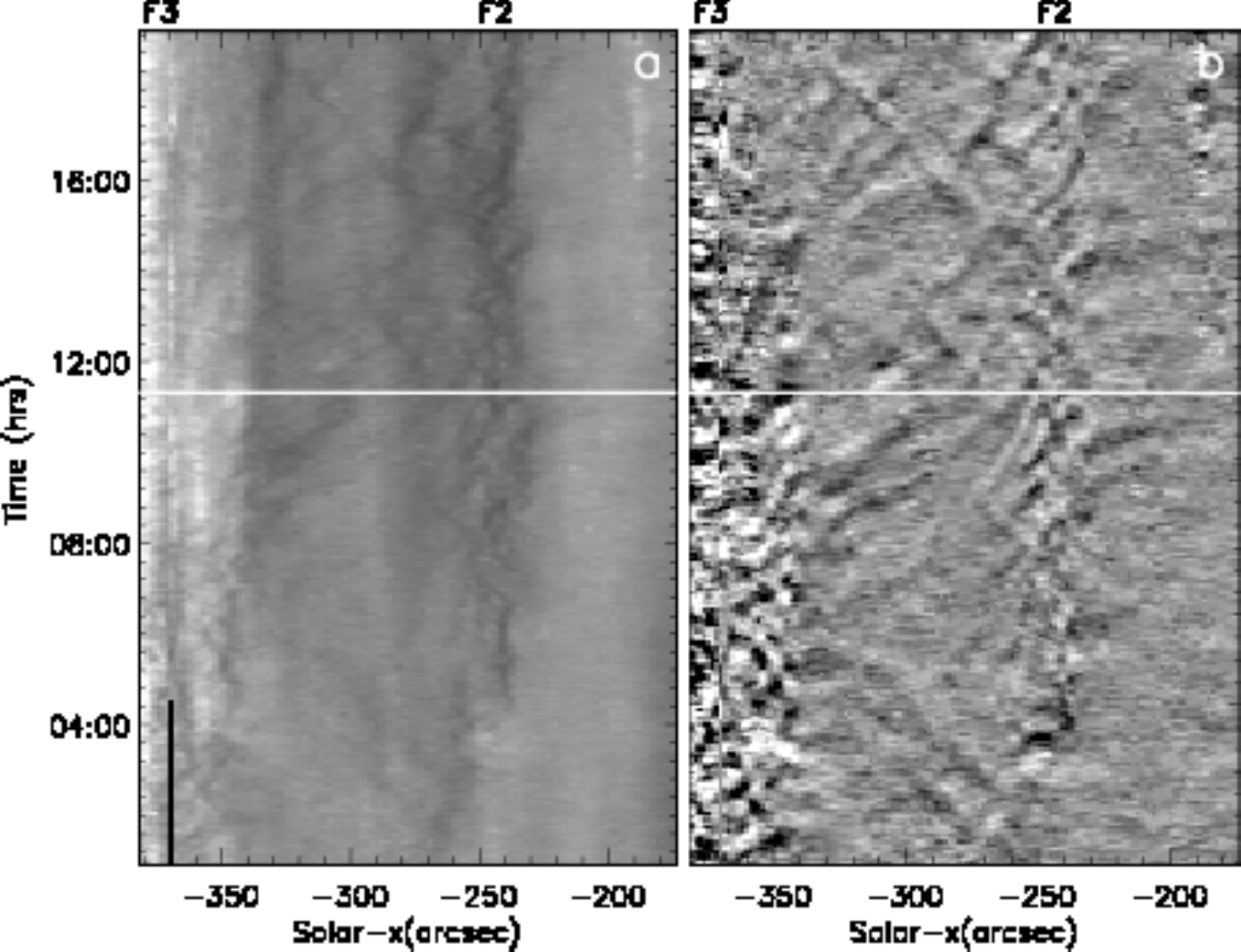}
      \caption{a) STEREO-A 195 \AA~ intensity time series and b) running difference time series along the dashed line shown in Figure~\ref{footpoints} on 13 February 2011. The
   white line represents the time (11:45 UT) when the spider-shaped prominence is observed.
   F2 and F3 are the positions of pillars.}
   \label{spi}
\end{figure}

\subsection{Footpoints of the prominence} 
\begin{figure*}
   \centering
   \includegraphics[width=\linewidth]{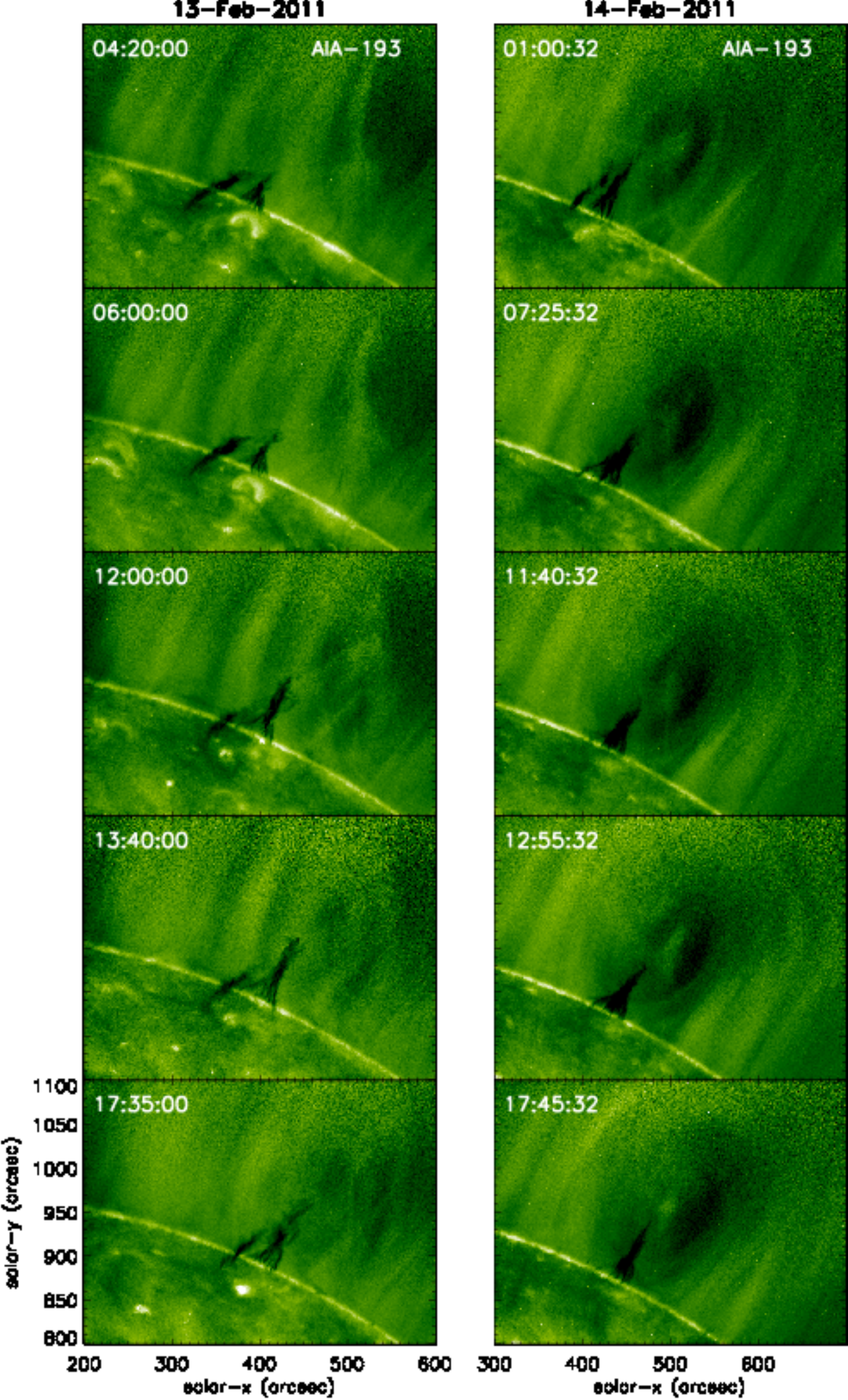}
   \caption{Evolution of the cavity system on 13 and 14 February seen in SDO/AIA 193 \AA\ images.}
\label{m193}
\end{figure*}

 In Figure~\ref{footpoints} STEREO-A 195 \AA\ and SDO 304 \AA/ 171 \AA\ composites taken at the same time on 13 February are shown.
In the SDO images cool plasma (brown) on the limb is often connected through the corona with warmer 171 \AA\ filaments (blue).
With the combination of two wavelengths
we can see the connections through the corona which form a sharp edge below the dark cavity above the
prominence. The dense pillars appear as dark structures in the 193 \AA\
images, taken at the same time and shown in Figure~\ref{m193}.
The connections between the pillars are not discernible in
the STEREO-A disk images or in the 193 \AA\ SDO limb images because they have a temperature of $\sim$ 0.6 MK \cite{par12}
where the \FeIX\ 171 \AA\ emission is strongest and the 193 \AA\ emission is due to \FeXII\ which is formed $\sim$ 2 MK.
The evolution of the prominence/filament system on 13 February can be followed in the
movie MOVIE13.

We derotated the STEREO images to 13 February at 00:05 UT, so that the SDO limb
position, shown as a meridional dashed line, moves on the STEREO image with time.
Identified features are indicated on Figure~\ref{footpoints}. Arrows on the images at 06:00~UT point to the three main pillars (F1, F2, F3)
and a bright point site, B1, that produced several brightenings.
The pillar, F1, is clearly visible at 06:00~UT in the STEREO images, but not in the 12:00~UT image.
The bright points, B2, B3, and B4, are indicated in the 12:00~UT images.

To investigate the on-disk dynamics, we created time series of the emission along a few horizontal slices in the STEREO images.
The positions of the time slices are shown as horizontal lines in the top-left frame of Figure~\ref{footpoints}. The long dot-dashed line cuts across
pillar F1 on the right, the filament channel, and the complicated section of
connected pillars (F2, F3, F4).
 In Figure~\ref{pillarcut}a and b, we show the time evolution of these features as seen in STEREO 195 \AA\
intensity and running-difference time-distance images.
In this representation, prominence pillars appear with a
criss-cross pattern as cold plasma moved into and out of the pillars.
The filament channel, which appears dark in the 195 \AA\ images,  shows almost no
variation except at its edge, B2, where there were continual small-scale brightenings.
There were several weak fronts coming from the bright point B2 and moving left towards
the direction of pillar F2. The red line indicates the position of one of the weak fronts which was moving
with a velocity of about 15~\kms. We note that there were no fronts moving in the other direction
from B2 into the filament channel, suggesting that these two regions were magnetically disconnected.

At F1 there is both the criss-cross pattern and continuous brightening until about 12:00~UT (see
Figure~\ref{pillarcut}). At the limb, there were continuous flows from 08:00 UT
until 12:00 UT. Subsequently, the prominence rotated behind the SDO limb, and
its lower portions near the photosphere were no longer visible.
The movie MOVIE13 shows plasma rising up from pillar F1 and abruptly stopping at the dark edge of a coronal cavity (Figure~\ref{footpoints} at 12:00~UT
image). The two-color images show that the flows are a combination of  171 and 304 \AA\ emitting plasma, with the chromospheric (304 \AA)
emission lower down and leading directly into the 171 \AA\ emission higher in the corona. Thus the
brightening at the footpoint of pillar F1 may be associated with the plasma heating and flow along a
 flux rope which magnetically connects F1 to F2. The change from 304 to 171 \AA\ emission can be due to
plasma heating along the flux tube or delayed ionization of previously heated plasma. This will be investigated in a later study
focussing on the thermal structure of the prominence plasma. The structure of the emission associated with the flows is
interesting because the prominence plasma must have stretched
 over about 150$\arcsec$ towards the east to connect to the pillar F2. This
 distance is about three times the line-of-sight distance travelled by the
 coronal flows seen in the spectroscopic observations of prominences analysed
 by \inlinecite{kuc03}.

 Additional flows from the top of the prominence out to the north started at 11:44 UT,
creating a spider-shaped prominence \cite{gil01}
with flows on either side of the main prominence pillar. Here the flows were going in one side (south) and out
the other (north).
  From the SDO images it looks as though the outflows were  triggered by the inflows
from the south. Possibly  a pressure increase from the inflowing plasma in the vicinity of a
flux-rope dip is sufficient to drive plasma out of the dip along a northern extension of the flux-rope. These flows were associated with
the generation of a new dark pillar in the STEREO images (in Figure~\ref{footpoints} marked as F4 on the 13:40~UT image).
 The STEREO movie (MOVIE13) shows at that this time there was a significant
amount of cold plasma falling from the corona towards pillar F4.
A time-series cut through this spider-shaped feature on the
STEREO images (horizontal dashed line in Figure~\ref{footpoints}) is shown in Figure~\ref{spi}.
The criss-cross flows after 11:30 UT (marked with a white line in Figure~\ref{spi})
are caused by the spider activity.

\begin{figure}
 \centering
 \includegraphics[width=\linewidth]{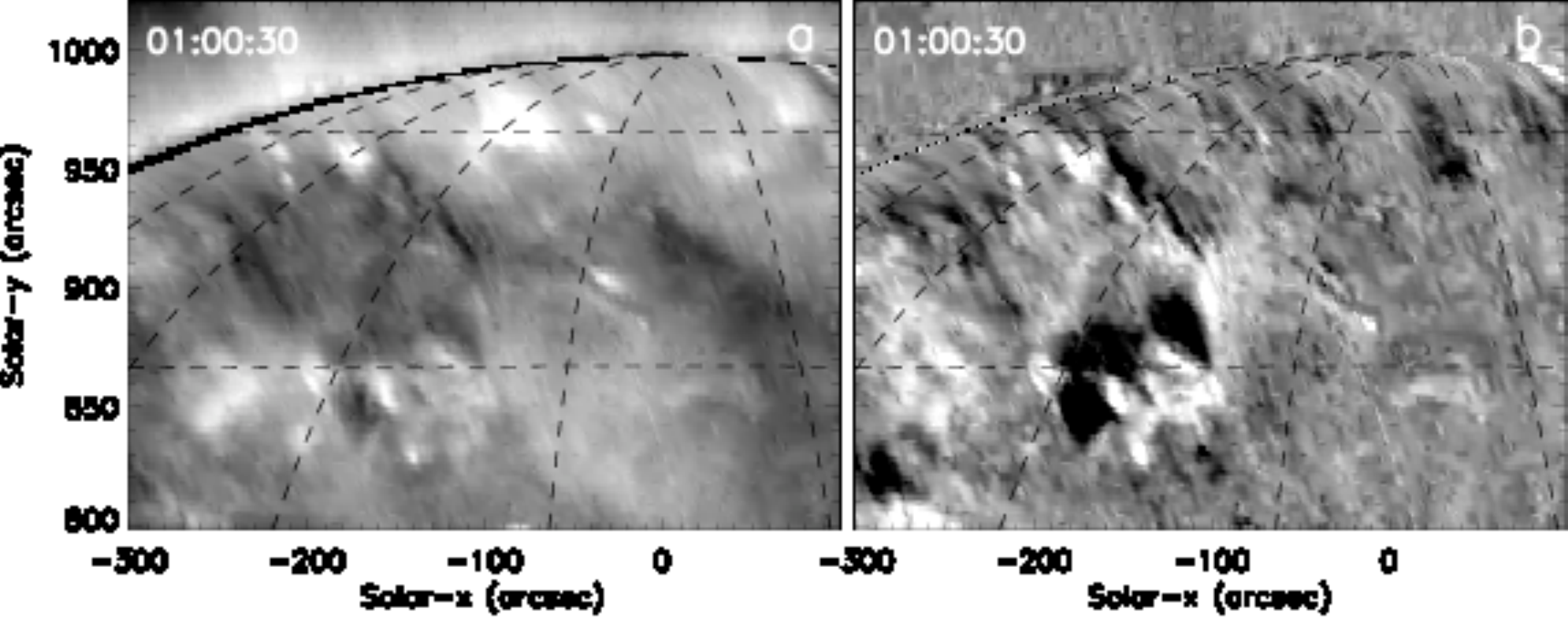}
\caption{Eruption on 14 February 01:00~UT. a) STEREO 195 \AA\ intensity image; b) STEREO 195 \AA\
base difference image overlaid with SDO grid lines.
The righthand longitude line shows the position of SDO limb. Grid spacing is $15^{\circ}$.
STEREO image has been derotated to 15 February at 00:05 UT.}
\label{eruptcut}
\end{figure}

\begin{figure}
 \centering
 \includegraphics[width=\linewidth]{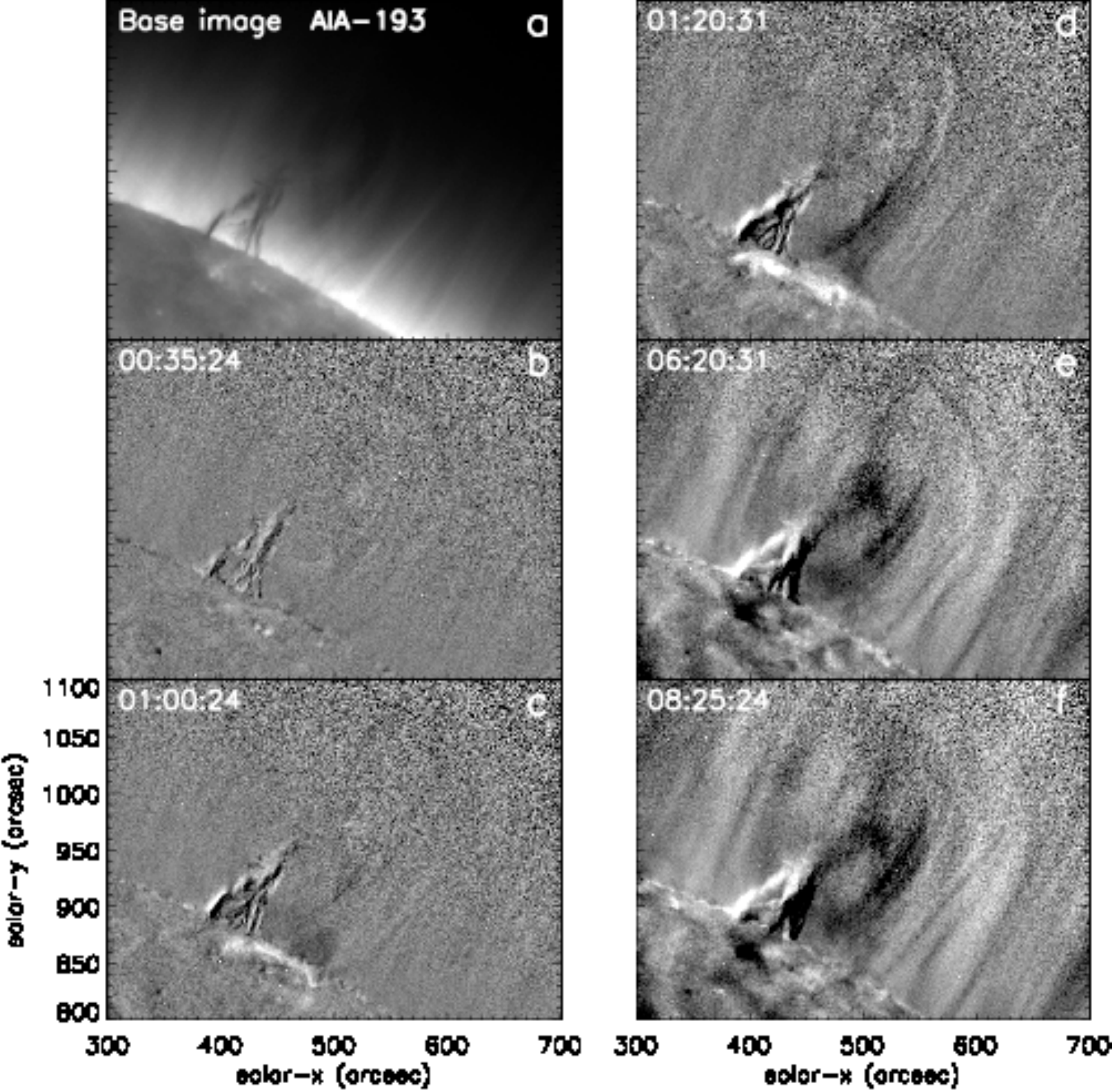}
\caption{Eruption on 14 February: (a) SDO 193 \AA\ base image, (b-f)
SDO 193 \AA\ base difference images show the dimming on either side of the prominence after the event at 01:00:24 UT.}
\label{basedif}
\end{figure}

\begin{figure*}
   \centering
   \includegraphics[width=\linewidth]{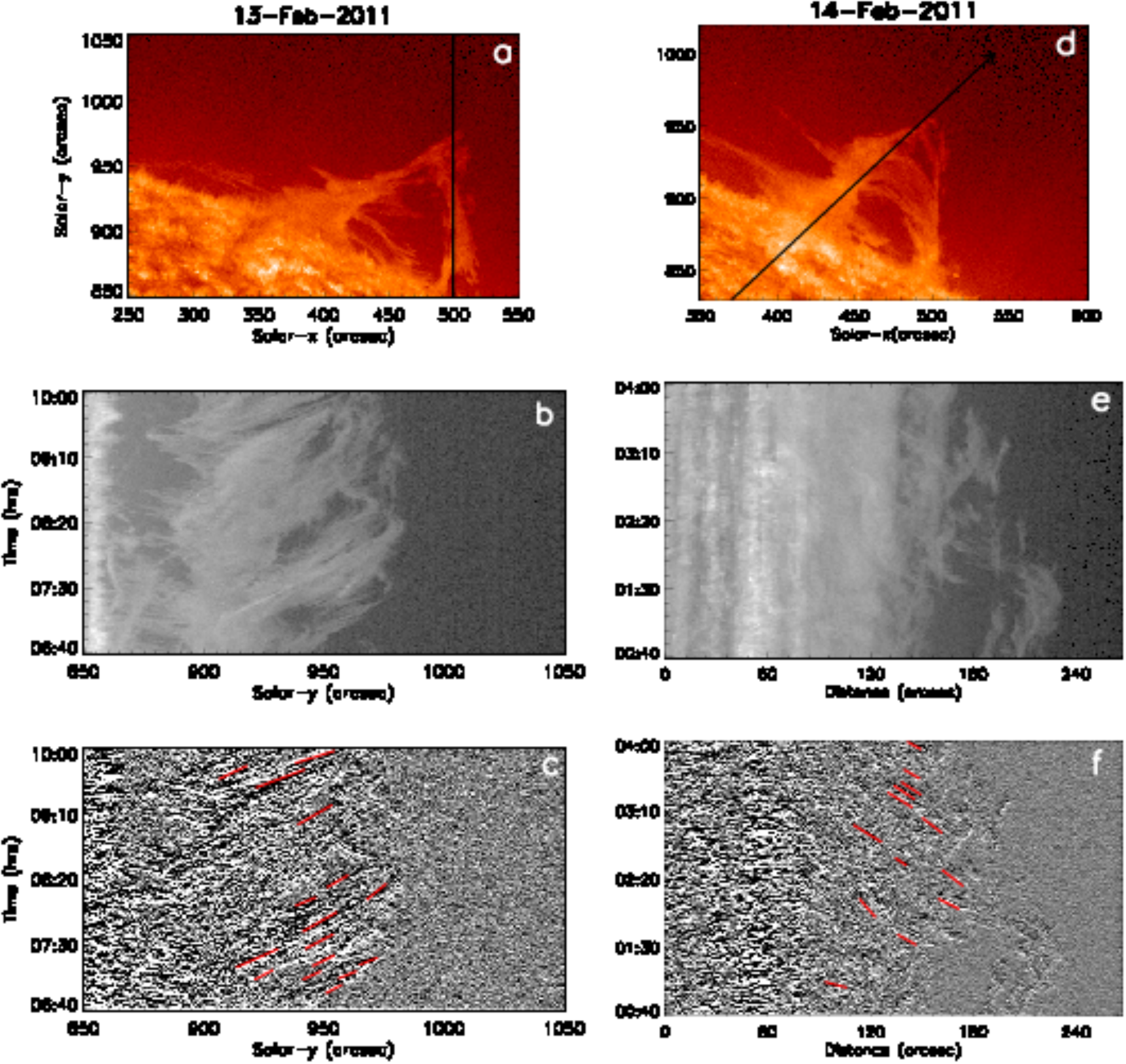}
      \caption{Prominence flows: a) SDO 304 \AA\ intensity image from 13 February at 08:20 UT. The vertical black line indicates
      the position of time series in (b) and (c); b) SDO 304 \AA\ intensity time
      series and c) running difference time series
      along the line in panel (a) on 13 February - red lines indicate the upflows; d) 304 \AA\ intensity image from 14 February at 01:40~UT.
      The arrow indicates the position of time series in (e) and (f); e) SDO 304 \AA\ intensity time series and
      f) running difference time series showing dominant downflow in pillars along the
      arrow in panel (d) - red lines indicate the downflows.}
      \label{upflows}
\end{figure*}

\subsection{ Event leading to cavity dimming} 
The evolution of the prominence on the 14~February is shown in
Figures~\ref{cavity} and \ref{m193}, and in the movie MOVIE14.
 During this day, two important things occurred. The first was a small
 event at B3 leading to coronal dimming over a significant portion of the filament channel, and the second was the
  breaking of the connection, F1-F2.

The event, shown in Figure~\ref{eruptcut}, occurred around 01:00~UT on 14 February.
 This was not a filament eruption but an event that had eruption-like features in on-disk images.
 It was associated with multiple brightenings seen in 304, 171 and 193 \AA, and
 jets of chromospheric material in the filament channel.  As seen from STEREO,
the event  produced a small EUV wave with a bright front, followed by a dimming
that swept over the footpoints of the prominence pillars, F2, F3, and F4.
The event can be clearly seen in the base-difference STEREO movie MOVIERD.
From SDO, the event was seen on the limb as a bright line along the prominence footpoints (Figure~\ref{cavity}, top right, B3)
 ejecting small jets of chromospheric plasma low down in the filament channel.
Although the event was not associated with a large change to the prominence structure, it caused
obvious dimming in the surrounding cavity as shown in the 193 \AA\ base-difference images
(Figure~\ref{basedif}). In the 171 \AA\ images
(Figure~\ref{cavity}), rather than the cavity itself being visible, we see
 enhanced emission from coronal plasma along its southern edge.
 The event and associated wave therefore heated the plasma inside the cavity and compressed the plasma along
its edge. It is possible that the event triggered a shift in the prominence flux rope as well, although this is not
visible in the image at 01:20 UT. The prominence shift seen in the 06:20 UT is due to the solar rotation.

When the connection between F1-F2 broke at 12:37 UT, a brightening appeared in the filament channel (see MOVIE14 and Figure~\ref{cavity}).
We find no signature of the brightening in the SDO 304, 171 \AA\ images
close to the limb (see Figure~\ref{cavity} at 12:55 UT). This indicates that the brightening was beyond the
limb (in SDO images) and occurred in the chromosphere.
It is interesting that when the connection between F1-F2 broke (see MOVIE14 and Figure~\ref{cavity}),
 the prominence pillar increased in height.

\subsection{Flow patterns in the prominence}
To analyse the flows in the prominence, we investigate the velocities
of the up- and downflows at different sites. Figure~\ref{upflows}a shows an SDO 304 \AA\ image when the  pillar F1 was prominent.
A black vertical line shows the position of the intensity and running difference time series shown in Figure~\ref{upflows}b,c. In Figure~\ref{upflows}b,
some of the stronger upflows are marked with red lines. Their average velocity is 15 $\pm\ $1.0 \kms. These upflows seem to have
been driven by heating at the footpoint of the pillar F1. We observe continuous  brightenings from the bright point
at the footpoint of F1 in the STEREO
195 \AA\ running-difference time-series image from 13
February (see Figure~\ref{pillarcut}). There is also an increase in the 171 to 304 \AA\ intensity ratio suggesting increasing temperature,
as the plasma rises from the photosphere to the corona (see movie MOVIE13).
According to the injection model \cite{wan99,cha03,Mackay10} heating may drive the plasma upward.
These strong upflows are easily seen in the SDO 304 \AA\ movie MOVIEUP.
Similar patterns of upflow have been observed on 14 February at pillar F1.

The central body of the prominence contained stable pillars of cool material, suspended
above the limb with moderate downflows.
We take a cut along the pillar F2 and F3 as shown in Figure~\ref{upflows}d.
Figure~\ref{upflows}e,f represents the SDO 304 \AA\
intensity and running difference time series from
14 February along the diagonal arrow shown in Figure~\ref{upflows}d.
The main flows in the pillars F2 and F3 were downward towards the solar surface. The average downflow velocity was 15.5 $\pm$ 1.5~\kms.
The dark streaks in the image represents the downflows highlighted with red lines.
These downflows seem to be the result of cool, condensed chromospheric
plasma that is falling down due to gravity \cite{dah98}. Small-scale upflows were also present but only for very short time periods.
Our estimations of velocities are in
agreement with that of \inlinecite{Berger08}. These velocities suggest that
prominence flows are moving approximately with the sound speed ($\sim$10~\kms) and less than the Alfv\'{e}n speed
($\sim$30~\kms). The observed velocities are consistent with \inlinecite{sch10}, who also reported Doppler motions of the
same magnitude, suggesting the presence of a magnetic-dip geometry.

\begin{figure*}
   \centering
   \includegraphics[width=\linewidth]{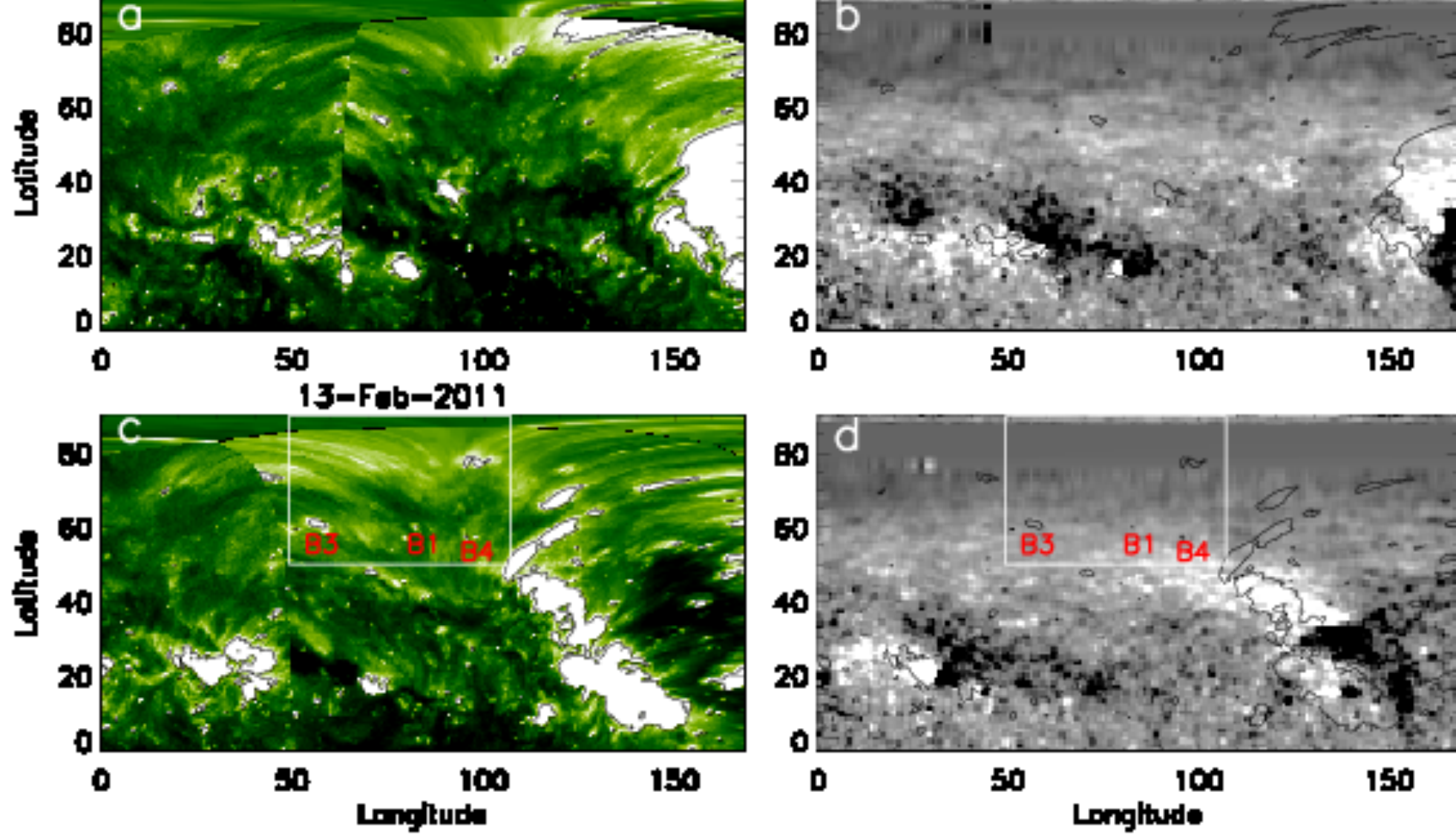}
      \caption{Magnetic field structure: a) Stoneyhurst projection of STEREO-A 195 \AA\
      intensity on 17 January 2011 with contours outlining the brightest patches;
      b) Gong synoptic map for 17 January 2011 with the same contours as in (a); c) as
      (a) one solar rotation after (13 February 2011); d) as (b) for 13 February 2011.
      Features identified in Figure~\ref{footpoints} and discussed in the text are marked.
      The white box outlines the region of the prominence/filament system is shown in Figure~\ref{footpoints}.}
      \label{carr_euv_lob}
\end{figure*}

\section{Discussion and Conclusions} 

\begin{figure}
   \centering
   \includegraphics[width=0.7\linewidth]{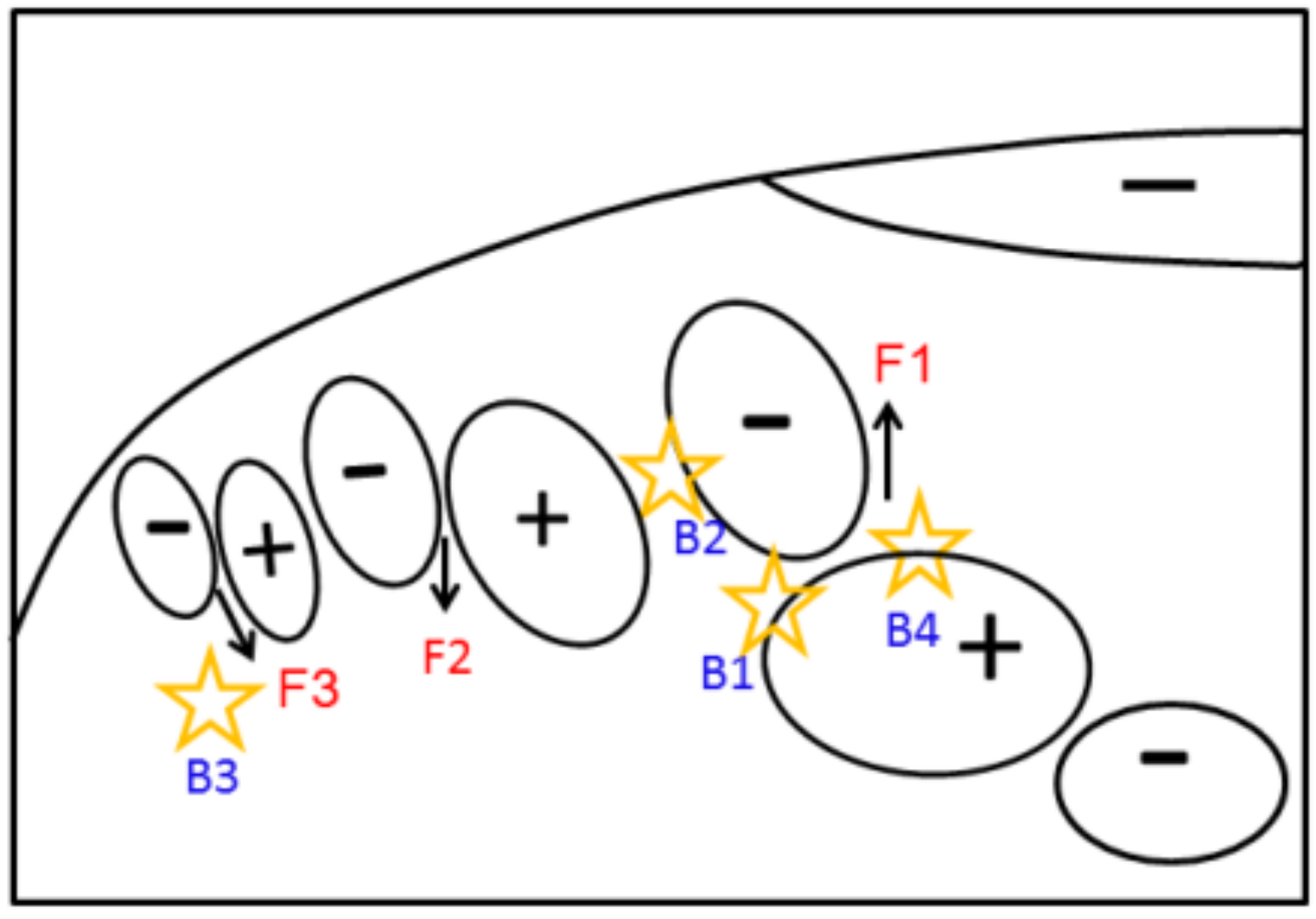}
      \caption{Sketch of our observations. Stars show sites of the various bright points.
      Pillars are marked as F1, F2 and F3. Positive and negative sign indicate magnetic polarity.}
      \label{skt}
\end{figure}

We have investigated the structure and dynamics of a polar crown prominence observed in the northern
hemisphere on 13--14 February 2011 seen simultaneously at the
limb from SDO and at disk center from STEREO-A. The observations were made during the
rising phase of the solar cycle when there was no obvious major polar coronal hole in either the SDO or the STEREO images of the northern pole.
We have identified three main pillars (vertical columns) of prominence plasma and four bright points which produced persistent brigtenings.
  Using time-series of STEREO on-disk images we deduce that the pillars are separated by hundreds of arcsecs and connected
by flows along the filament channel. The character of the start and end sites are determined
using running difference time-slice images of on-disk emission which show distinctly different behaviour for bright points,
flows, and pillars.
 During the two days studied, we noticed that microflaring (at the footpoint of pillar F1) produced 171 \AA\
prominence flows that rise up and stop abruptly at a sharp concave edge in the
corona because the plasma is constrained to flow along the flux rope.

One of the motivations for this study was to observe and study the dynamics of a prominence-cavity system.
 Initially, the cavity was visible as a concave, bowl-shaped dark region in 171 \AA\ emission  above the prominence. It was particularly noticeable
when there were flows between pillars F1 and F2. When the connection between the two main pillars, F1 and F2,
broke,  the magnetic field in the
prominence/filament system reconfigured and the prominence increased in height.

The dark cavity region seen in SDO coincided with a dark region close to
the filament channel, in the STEREO disk images.
Unlike regions of the channel with dense, prominence plasma,
this darker region did not show signs of flows in the STEREO time-series. We therefore predict that the darkness was
due to its low density, rather than cool prominence plasma and that it is the disk signature of the prominence cavity.
We found a similar region of reduced 193 \AA\ emission at approximately the same position in the SDO images when the
prominence/filament channel was on the disk from SDO.
In both the STEREO and the disk SDO images, persistent brightening was seen along the edge of this region
of reduced 195/193 \AA\ emission.

The presence of the cavity also became apparent after a small flare-like brightening and EUV wave inside the
filament channel at 01:00 UT on 14 February.
This caused a coronal dimming that was immediately visible in the STEREO images and then later as a bowl-shaped dark region in 171 \AA\ emission
 (e.g. \opencite{gib06}) in the limb EUV images. It remained as a visible cavity in 171 \AA\
until the region rotated around the limb.

 We therefore find that the presence of the cavity was made visible in the EUV images in two circumstances.
The first was flows along the prominence on 13 February that highlighted the boundary between the prominence
and the cavity and the second was due to heating and compression
caused by an eruption-like event close to the filament
channel on 14 February.

To understand the prominence/filament structures in relation to the large-scale magnetic field, we compared
the sites of bright points and pillars to the underlying photospheric flux. For this purpose we
used line-of-sight magnetograms from GONG and composite SDO and STEREO-A 195 \AA\ EUV images.
Figure~\ref{carr_euv_lob} shows EUV images and contours of the images overlaid on the GONG
magnetic fields for the 13 February 2011. The interpretation of these images at such high latitudes is
ambiguous but the large scale is still discernible.
To the south west was strong
positive polarity active region flux that has diffused to this position during the previous 27
days (Figure~\ref{carr_euv_lob}a,b).
A close-up of the diffused positive flux and higher latitude negative flux is shown in the HMI image (Figure~\ref{hmi}a).
The cancellation of the newly diffused positive flux with pre-existing negative flux elements
 is the reason for the
many microflares at B1, B3 and B4 (Figure~\ref{carr_euv_lob}).

A sketch of the inferred magnetic topology and sites of the prominence features is shown in Figure~\ref{skt}. Since the
observations were made in the rising phase of the solar cycle when
the negative polarity is dominant over the northern pole, we have placed a region of negative field over the pole.
The newly diffused active region is on the south west.
The positions of the other magnetic flux regions are more speculative and are chosen to reflect the
structure seen in the GONG map and the observation that filaments form between adjacent active regions.
\inlinecite{Martens01} proposed a flux linkage model to explain the chains
of polar crown filaments. Here we have the situation where the arrival of new active region flux is merging
with an already formed chain due to magnetic field reconnection at B2, B1, and
 the footpoint of pillar F1 which led to the injection of flux and plasma into the prominence.

In this paper we have demonstrated the possibility of being able to dissect polar crown prominences into individual pillars, and
filaments flows using observations from spacecraft separated by about 90\degree.
 These dynamics suggest the presence of complex magnetic geometries
in the filament channel. The arrival of diffused active region flux seems
to play an important role in the evolution of this prominence/filament
system.

\begin{acks}
 We would like to thank the referee for his/her constructive comments.
SDO data are courtesy of the NASA/SDO AIA
and HMI science teams. STEREO data are courtesy
of the STEREO Sun Earth Connection Coronal and Heliospheric Investigation (SECCHI)
team. GONG magnetogram data obtained by the Global Oscillation Network Group (GONG) project,
managed by the National Solar Observatory, which is operated by AURA, Inc. under a cooperative agreement with the National Science Foundation.
NKP acknowledges the facilities provided by the MPS. One of the authors, SKT, is supported by an appointment
to the NASA Postdoctoral Program at
the NASA Marshall Space Flight Center, administered by Oak
Ridge Associated Universities through a contract with NASA.
\end{acks}

\bibliographystyle{spr-mp-sola}

\begin{thebibliography}{0}
\ifx \bisbn   \undefined \def \bisbn  #1{ISBN #1}\fi
\ifx \binits  \undefined \def \binits#1{#1}\fi
\ifx \bauthor  \undefined \def \bauthor#1{#1}\fi
\ifx \batitle  \undefined \def \batitle#1{#1}\fi
\ifx \bjtitle  \undefined \def \bjtitle#1{\textit{#1}}\fi
\ifx \bvolume  \undefined \def \bvolume#1{\textbf{#1}}\fi
\ifx \byear  \undefined \def \byear#1{#1}\fi
\ifx \bissue  \undefined \def \bissue#1{#1}\fi
\ifx \bfpage  \undefined \def \bfpage#1{#1}\fi
\ifx \blpage  \undefined \def \blpage #1{#1}\fi
\ifx \burl  \undefined \def \burl#1{\textsf{#1}}\fi
\ifx \href  \undefined \def \href#1#2{\textsf{#2}}\fi
\ifx \doiurl  \undefined \def
  \doiurl#1{\href{http://dx.doi.org/#1}{\textsf{#1}}}\fi
\ifx \betal  \undefined \def \betal{\textit{et al.}}\fi
\ifx \binstitute  \undefined \def \binstitute#1{#1}\fi
\ifx \bctitle  \undefined \def \bctitle#1{#1}\fi
\ifx \beditor  \undefined \def \beditor#1{#1}\fi
\ifx \bpublisher  \undefined \def \bpublisher#1{#1}\fi
\ifx \bbtitle  \undefined \def \bbtitle#1{\textit{#1}}\fi
\ifx \bedition  \undefined \def \bedition#1{#1}\fi
\ifx \bseriesno  \undefined \def \bseriesno#1{\textbf{#1}}\fi
\ifx \blocation  \undefined \def \blocation#1{#1}\fi
\ifx \bsertitle  \undefined \def \bsertitle#1{\textit{#1}}\fi
\ifx \bsnm \undefined \def \bsnm#1{#1}\fi
\ifx \bsuffix \undefined \def \bsuffix#1{#1}\fi
\ifx \bparticle \undefined \def \bparticle#1{#1}\fi
\ifx \barticle \undefined \def \barticle{}\fi
\ifx \botherref \undefined \def \botherref{}\fi
\ifx \url \undefined \def \url#1{\textsf{#1}}\fi
\ifx \bchapter \undefined \def \bchapter{}\fi
\ifx \bbook \undefined \def \bbook{}\fi
\ifx \bcomment \undefined \def \bcomment#1{#1}\fi
\ifx \oauthor \undefined \def \oauthor#1{#1}\fi
\ifx \citeauthoryear \undefined \def \citeauthoryear#1{#1}\fi
\def \endbibitem {}

\end{thebibliography}


\begin{thebibliography}{66}
\ifx \bisbn   \undefined \def \bisbn  #1{ISBN #1}\fi
\ifx \binits  \undefined \def \binits#1{#1}\fi
\ifx \bauthor  \undefined \def \bauthor#1{#1}\fi
\ifx \batitle  \undefined \def \batitle#1{#1}\fi
\ifx \bjtitle  \undefined \def \bjtitle#1{\textit{#1}}\fi
\ifx \bvolume  \undefined \def \bvolume#1{\textbf{#1}}\fi
\ifx \byear  \undefined \def \byear#1{#1}\fi
\ifx \bissue  \undefined \def \bissue#1{#1}\fi
\ifx \bfpage  \undefined \def \bfpage#1{#1}\fi
\ifx \blpage  \undefined \def \blpage #1{#1}\fi
\ifx \burl  \undefined \def \burl#1{\textsf{#1}}\fi
\ifx \href  \undefined \def \href#1#2{\textsf{#2}}\fi
\ifx \doiurl  \undefined \def
  \doiurl#1{\href{http://dx.doi.org/#1}{\textsf{#1}}}\fi
\ifx \betal  \undefined \def \betal{\textit{et al.}}\fi
\ifx \binstitute  \undefined \def \binstitute#1{#1}\fi
\ifx \bctitle  \undefined \def \bctitle#1{#1}\fi
\ifx \beditor  \undefined \def \beditor#1{#1}\fi
\ifx \bpublisher  \undefined \def \bpublisher#1{#1}\fi
\ifx \bbtitle  \undefined \def \bbtitle#1{\textit{#1}}\fi
\ifx \bedition  \undefined \def \bedition#1{#1}\fi
\ifx \bseriesno  \undefined \def \bseriesno#1{\textbf{#1}}\fi
\ifx \blocation  \undefined \def \blocation#1{#1}\fi
\ifx \bsertitle  \undefined \def \bsertitle#1{\textit{#1}}\fi
\ifx \bsnm \undefined \def \bsnm#1{#1}\fi
\ifx \bsuffix \undefined \def \bsuffix#1{#1}\fi
\ifx \bparticle \undefined \def \bparticle#1{#1}\fi
\ifx \barticle \undefined \def \barticle{}\fi
\ifx \botherref \undefined \def \botherref{}\fi
\ifx \url \undefined \def \url#1{\textsf{#1}}\fi
\ifx \bchapter \undefined \def \bchapter{}\fi
\ifx \bbook \undefined \def \bbook{}\fi
\ifx \bcomment \undefined \def \bcomment#1{#1}\fi
\ifx \oauthor \undefined \def \oauthor#1{#1}\fi
\ifx \citeauthoryear \undefined \def \citeauthoryear#1{#1}\fi
\def \endbibitem {}

\bibitem[\protect\citeauthoryear{{Antiochos}, {Dahlburg}, and
  {Klimchuk}}{1994}]{ant94}
\begin{barticle}
\bauthor{\bsnm{{Antiochos}}, \binits{S.K.}}, \bauthor{\bsnm{{Dahlburg}},
  \binits{R.B.}}, \bauthor{\bsnm{{Klimchuk}}, \binits{J.A.}}:
\byear{1994},
\batitle{{The magnetic field of solar prominences}}.
\bjtitle{\apjl}
\bvolume{420},
\bfpage{L41}\,--\,\blpage{L44}.
doi:\doiurl{10.1086/187158}.
\end{barticle}
\endbibitem

\bibitem[\protect\citeauthoryear{{Anzer}}{1994}]{anz94}
\begin{botherref}
\oauthor{\bsnm{{Anzer}}, \binits{U.}}:
1994,
{Remarks on two-dimensional magnetic arcade models of coronal structures.}
In: {Rusin}, V., {Heinzel}, P., {Vial}, J.C. (eds.)
\textit{IAU Colloq. 144: Solar Coronal Structures},
309\,--\,314.
\end{botherref}
\endbibitem

\bibitem[\protect\citeauthoryear{{Anzer} and {Heinzel}}{2005}]{anz05}
\begin{barticle}
\bauthor{\bsnm{{Anzer}}, \binits{U.}}, \bauthor{\bsnm{{Heinzel}}, \binits{P.}}:
\byear{2005},
\batitle{{On the Nature of Dark Extreme Ultraviolet Structures Seen by SOHO/EIT
  and TRACE}}.
\bjtitle{\apj}
\bvolume{622},
\bfpage{714}\,--\,\blpage{721}.
doi:\doiurl{10.1086/427817}.
\end{barticle}
\endbibitem

\bibitem[\protect\citeauthoryear{{Aulanier} and {D\'{e}moulin}}{1998}]{aul98}
\begin{barticle}
\bauthor{\bsnm{{Aulanier}}, \binits{G.}}, \bauthor{\bsnm{{D\'{e}moulin}},
  \binits{P.}}:
\byear{1998},
\batitle{{3-D magnetic configurations supporting prominences. I. The natural
  presence of lateral feet}}.
\bjtitle{\aap}
\bvolume{329},
\bfpage{1125}\,--\,\blpage{1137}.
\end{barticle}
\endbibitem

\bibitem[\protect\citeauthoryear{{Aulanier}, {DeVore}, and
  {Antiochos}}{2002}]{aul02}
\begin{barticle}
\bauthor{\bsnm{{Aulanier}}, \binits{G.}}, \bauthor{\bsnm{{DeVore}},
  \binits{C.R.}}, \bauthor{\bsnm{{Antiochos}}, \binits{S.K.}}:
\byear{2002},
\batitle{{Prominence Magnetic Dips in Three-Dimensional Sheared Arcades}}.
\bjtitle{\apjl}
\bvolume{567},
\bfpage{L97}\,--\,\blpage{L101}.
doi:\doiurl{10.1086/339436}.
\end{barticle}
\endbibitem

\bibitem[\protect\citeauthoryear{{B\c{a}k-St{\c e}{\'s}licka}
  \textit{et~al.}}{2013}]{bak13}
\begin{barticle}
\bauthor{\bsnm{{B\c{a}k-St{\c e}{\'s}licka}}, \binits{U.}},
  \bauthor{\bsnm{{Gibson}}, \binits{S.E.}}, \bauthor{\bsnm{{Fan}},
  \binits{Y.}}, \bauthor{\bsnm{{Bethge}}, \binits{C.}},
  \bauthor{\bsnm{{Forland}}, \binits{B.}}, \bauthor{\bsnm{{Rachmeler}},
  \binits{L.A.}}:
\byear{2013},
\batitle{{The Magnetic Structure of Solar Prominence Cavities: New
  Observational Signature Revealed by Coronal Magnetometry}}.
\bjtitle{\apjl}
\bvolume{770},
\bfpage{L28}.
doi:\doiurl{10.1088/2041-8205/770/2/L28}.
\end{barticle}
\endbibitem

\bibitem[\protect\citeauthoryear{{Berger}, {Liu}, and {Low}}{2012}]{berger12}
\begin{barticle}
\bauthor{\bsnm{{Berger}}, \binits{T.E.}}, \bauthor{\bsnm{{Liu}}, \binits{W.}},
  \bauthor{\bsnm{{Low}}, \binits{B.C.}}:
\byear{2012},
\batitle{{SDO/AIA Detection of Solar Prominence Formation within a Coronal
  Cavity}}.
\bjtitle{\apjl}
\bvolume{758},
\bfpage{L37}.
doi:\doiurl{10.1088/2041-8205/758/2/L37}.
\end{barticle}
\endbibitem

\bibitem[\protect\citeauthoryear{{Berger} \textit{et~al.}}{2008}]{Berger08}
\begin{barticle}
\bauthor{\bsnm{{Berger}}, \binits{T.E.}}, \bauthor{\bsnm{{Shine}},
  \binits{R.A.}}, \bauthor{\bsnm{{Slater}}, \binits{G.L.}},
  \bauthor{\bsnm{{Tarbell}}, \binits{T.D.}}, \bauthor{\bsnm{{Title}},
  \binits{A.M.}}, \bauthor{\bsnm{{Okamoto}}, \binits{T.J.}},
  \bauthor{\bsnm{{Ichimoto}}, \binits{K.}}, \bauthor{\bsnm{{Katsukawa}},
  \binits{Y.}}, \bauthor{\bsnm{{Suematsu}}, \binits{Y.}},
  \bauthor{\bsnm{{Tsuneta}}, \binits{S.}}, \bauthor{\bsnm{{Lites}},
  \binits{B.W.}}, \bauthor{\bsnm{{Shimizu}}, \binits{T.}}:
\byear{2008},
\batitle{{Hinode SOT Observations of Solar Quiescent Prominence Dynamics}}.
\bjtitle{\apjl}
\bvolume{676},
\bfpage{L89}\,--\,\blpage{L92}.
doi:\doiurl{10.1086/587171}.
\end{barticle}
\endbibitem

\bibitem[\protect\citeauthoryear{{Berger} \textit{et~al.}}{2011}]{Berger11}
\begin{barticle}
\bauthor{\bsnm{{Berger}}, \binits{T.}}, \bauthor{\bsnm{{Testa}}, \binits{P.}},
  \bauthor{\bsnm{{Hillier}}, \binits{A.}}, \bauthor{\bsnm{{Boerner}},
  \binits{P.}}, \bauthor{\bsnm{{Low}}, \binits{B.C.}},
  \bauthor{\bsnm{{Shibata}}, \binits{K.}}, \bauthor{\bsnm{{Schrijver}},
  \binits{C.}}, \bauthor{\bsnm{{Tarbell}}, \binits{T.}},
  \bauthor{\bsnm{{Title}}, \binits{A.}}:
\byear{2011},
\batitle{{Magneto-thermal convection in solar prominences}}.
\bjtitle{\nat}
\bvolume{472},
\bfpage{197}\,--\,\blpage{200}.
doi:\doiurl{10.1038/nature09925}.
\end{barticle}
\endbibitem

\bibitem[\protect\citeauthoryear{{Chae}}{2003}]{cha03}
\begin{barticle}
\bauthor{\bsnm{{Chae}}, \binits{J.}}:
\byear{2003},
\batitle{{The Formation of a Prominence in NOAA Active Region 8668. II. Trace
  Observations of Jets and Eruptions Associated with Canceling Magnetic
  Features}}.
\bjtitle{\apj}
\bvolume{584},
\bfpage{1084}\,--\,\blpage{1094}.
doi:\doiurl{10.1086/345739}.
\end{barticle}
\endbibitem

\bibitem[\protect\citeauthoryear{{Chae} \textit{et~al.}}{2008}]{cha08}
\begin{barticle}
\bauthor{\bsnm{{Chae}}, \binits{J.}}, \bauthor{\bsnm{{Ahn}}, \binits{K.}},
  \bauthor{\bsnm{{Lim}}, \binits{E.K.}}, \bauthor{\bsnm{{Choe}},
  \binits{G.S.}}, \bauthor{\bsnm{{Sakurai}}, \binits{T.}}:
\byear{2008},
\batitle{{Persistent Horizontal Flows and Magnetic Support of Vertical Threads
  in a Quiescent Prominence}}.
\bjtitle{\apjl}
\bvolume{689},
\bfpage{L73}\,--\,\blpage{L76}.
doi:\doiurl{10.1086/595785}.
\end{barticle}
\endbibitem

\bibitem[\protect\citeauthoryear{{Dahlburg}, {Antiochos}, and
  {Klimchuk}}{1998}]{dah98}
\begin{barticle}
\bauthor{\bsnm{{Dahlburg}}, \binits{R.B.}}, \bauthor{\bsnm{{Antiochos}},
  \binits{S.K.}}, \bauthor{\bsnm{{Klimchuk}}, \binits{J.A.}}:
\byear{1998},
\batitle{{Prominence Formation by Localized Heating}}.
\bjtitle{\apj}
\bvolume{495},
\bfpage{485}\,--\,\blpage{490}.
doi:\doiurl{10.1086/305286}.
\end{barticle}
\endbibitem

\bibitem[\protect\citeauthoryear{{Del Zanna}, {O'Dwyer}, and
  {Mason}}{2011}]{del11}
\begin{barticle}
\bauthor{\bsnm{{Del Zanna}}, \binits{G.}}, \bauthor{\bsnm{{O'Dwyer}},
  \binits{B.}}, \bauthor{\bsnm{{Mason}}, \binits{H.E.}}:
\byear{2011},
\batitle{{SDO AIA and Hinode EIS observations of ''warm'' loops}}.
\bjtitle{\aap}
\bvolume{535},
\bfpage{A46}.
doi:\doiurl{10.1051/0004-6361/201117470}.
\end{barticle}
\endbibitem

\bibitem[\protect\citeauthoryear{{DeVore} and {Antiochos}}{2000}]{dev00}
\begin{barticle}
\bauthor{\bsnm{{DeVore}}, \binits{C.R.}}, \bauthor{\bsnm{{Antiochos}},
  \binits{S.K.}}:
\byear{2000},
\batitle{{Dynamical Formation and Stability of Helical Prominence Magnetic
  Fields}}.
\bjtitle{\apj}
\bvolume{539},
\bfpage{954}\,--\,\blpage{963}.
doi:\doiurl{10.1086/309275}.
\end{barticle}
\endbibitem

\bibitem[\protect\citeauthoryear{{Dud{\'{\i}}k} \textit{et~al.}}{2012}]{dud12}
\begin{barticle}
\bauthor{\bsnm{{Dud{\'{\i}}k}}, \binits{J.}}, \bauthor{\bsnm{{Aulanier}},
  \binits{G.}}, \bauthor{\bsnm{{Schmieder}}, \binits{B.}},
  \bauthor{\bsnm{{Zapi{\'o}r}}, \binits{M.}}, \bauthor{\bsnm{{Heinzel}},
  \binits{P.}}:
\byear{2012},
\batitle{{Magnetic Topology of Bubbles in Quiescent Prominences}}.
\bjtitle{\apj}
\bvolume{761},
\bfpage{9}.
doi:\doiurl{10.1088/0004-637X/761/1/9}.
\end{barticle}
\endbibitem

\bibitem[\protect\citeauthoryear{{Gaizauskas} \textit{et~al.}}{1997}]{gai97}
\begin{barticle}
\bauthor{\bsnm{{Gaizauskas}}, \binits{V.}}, \bauthor{\bsnm{{Zirker}},
  \binits{J.B.}}, \bauthor{\bsnm{{Sweetland}}, \binits{C.}},
  \bauthor{\bsnm{{Kovacs}}, \binits{A.}}:
\byear{1997},
\batitle{{Formation of a Solar Filament Channel}}.
\bjtitle{\apj}
\bvolume{479},
\bfpage{448}\,--\,\blpage{457}.
doi:\doiurl{10.1086/512788}.
\end{barticle}
\endbibitem

\bibitem[\protect\citeauthoryear{{Gibson} and {Fan}}{2006}]{gib06}
\begin{barticle}
\bauthor{\bsnm{{Gibson}}, \binits{S.E.}}, \bauthor{\bsnm{{Fan}}, \binits{Y.}}:
\byear{2006},
\batitle{{Coronal prominence structure and dynamics: A magnetic flux rope
  interpretation}}.
\bjtitle{\jgr}
\bvolume{111},
\bfpage{12103}.
doi:\doiurl{10.1029/2006JA011871}.
\end{barticle}
\endbibitem

\bibitem[\protect\citeauthoryear{{Gibson} \textit{et~al.}}{2006}]{Gibson06}
\begin{barticle}
\bauthor{\bsnm{{Gibson}}, \binits{S.E.}}, \bauthor{\bsnm{{Foster}},
  \binits{D.}}, \bauthor{\bsnm{{Burkepile}}, \binits{J.}}, \bauthor{\bsnm{{de
  Toma}}, \binits{G.}}, \bauthor{\bsnm{{Stanger}}, \binits{A.}}:
\byear{2006},
\batitle{{The Calm before the Storm: The Link between Quiescent Cavities and
  Coronal Mass Ejections}}.
\bjtitle{\apj}
\bvolume{641},
\bfpage{590}\,--\,\blpage{605}.
doi:\doiurl{10.1086/500446}.
\end{barticle}
\endbibitem

\bibitem[\protect\citeauthoryear{{Gibson} \textit{et~al.}}{2010}]{gib10}
\begin{barticle}
\bauthor{\bsnm{{Gibson}}, \binits{S.E.}}, \bauthor{\bsnm{{Kucera}},
  \binits{T.A.}}, \bauthor{\bsnm{{Rastawicki}}, \binits{D.}},
  \bauthor{\bsnm{{Dove}}, \binits{J.}}, \bauthor{\bsnm{{de Toma}},
  \binits{G.}}, \bauthor{\bsnm{{Hao}}, \binits{J.}}, \bauthor{\bsnm{{Hill}},
  \binits{S.}}, \bauthor{\bsnm{{Hudson}}, \binits{H.S.}},
  \bauthor{\bsnm{{Marqu{\'e}}}, \binits{C.}}, \bauthor{\bsnm{{McIntosh}},
  \binits{P.S.}}, \bauthor{\bsnm{{Rachmeler}}, \binits{L.}},
  \bauthor{\bsnm{{Reeves}}, \binits{K.K.}}, \bauthor{\bsnm{{Schmieder}},
  \binits{B.}}, \bauthor{\bsnm{{Schmit}}, \binits{D.J.}},
  \bauthor{\bsnm{{Seaton}}, \binits{D.B.}}, \bauthor{\bsnm{{Sterling}},
  \binits{A.C.}}, \bauthor{\bsnm{{Tripathi}}, \binits{D.}},
  \bauthor{\bsnm{{Williams}}, \binits{D.R.}}, \bauthor{\bsnm{{Zhang}},
  \binits{M.}}:
\byear{2010},
\batitle{{Three-dimensional Morphology of a Coronal Prominence Cavity}}.
\bjtitle{\apj}
\bvolume{724},
\bfpage{1133}\,--\,\blpage{1146}.
doi:\doiurl{10.1088/0004-637X/724/2/1133}.
\end{barticle}
\endbibitem

\bibitem[\protect\citeauthoryear{{Gilbert}, {Holzer}, and
  {Burkepile}}{2001}]{gil01}
\begin{barticle}
\bauthor{\bsnm{{Gilbert}}, \binits{H.R.}}, \bauthor{\bsnm{{Holzer}},
  \binits{T.E.}}, \bauthor{\bsnm{{Burkepile}}, \binits{J.T.}}:
\byear{2001},
\batitle{{Observational Interpretation of an Active Prominence on 1999 May 1}}.
\bjtitle{\apj}
\bvolume{549},
\bfpage{1221}\,--\,\blpage{1230}.
doi:\doiurl{10.1086/319444}.
\end{barticle}
\endbibitem

\bibitem[\protect\citeauthoryear{{Habbal} \textit{et~al.}}{2010}]{hab10}
\begin{barticle}
\bauthor{\bsnm{{Habbal}}, \binits{S.R.}}, \bauthor{\bsnm{{Druckm{\"u}ller}},
  \binits{M.}}, \bauthor{\bsnm{{Morgan}}, \binits{H.}},
  \bauthor{\bsnm{{Scholl}}, \binits{I.}}, \bauthor{\bsnm{{Ru{\v s}in}},
  \binits{V.}}, \bauthor{\bsnm{{Daw}}, \binits{A.}}, \bauthor{\bsnm{{Johnson}},
  \binits{J.}}, \bauthor{\bsnm{{Arndt}}, \binits{M.}}:
\byear{2010},
\batitle{{Total Solar Eclipse Observations of Hot Prominence Shrouds}}.
\bjtitle{\apj}
\bvolume{719},
\bfpage{1362}\,--\,\blpage{1369}.
doi:\doiurl{10.1088/0004-637X/719/2/1362}.
\end{barticle}
\endbibitem

\bibitem[\protect\citeauthoryear{{Harvey} \textit{et~al.}}{1996}]{harvey96}
\begin{barticle}
\bauthor{\bsnm{{Harvey}}, \binits{J.W.}}, \bauthor{\bsnm{{Hill}}, \binits{F.}},
  \bauthor{\bsnm{{Hubbard}}, \binits{R.P.}}, \bauthor{\bsnm{{Kennedy}},
  \binits{J.R.}}, \bauthor{\bsnm{{Leibacher}}, \binits{J.W.}},
  \bauthor{\bsnm{{Pintar}}, \binits{J.A.}}, \bauthor{\bsnm{{Gilman}},
  \binits{P.A.}}, \bauthor{\bsnm{{Noyes}}, \binits{R.W.}},
  \bauthor{\bsnm{{Title}}, \binits{A.M.}}, \bauthor{\bsnm{{Toomre}},
  \binits{J.}}, \bauthor{\bsnm{{Ulrich}}, \binits{R.K.}},
  \bauthor{\bsnm{{Bhatnagar}}, \binits{A.}}, \bauthor{\bsnm{{Kennewell}},
  \binits{J.A.}}, \bauthor{\bsnm{{Marquette}}, \binits{W.}},
  \bauthor{\bsnm{{Patron}}, \binits{J.}}, \bauthor{\bsnm{{Saa}}, \binits{O.}},
  \bauthor{\bsnm{{Yasukawa}}, \binits{E.}}:
\byear{1996},
\batitle{{The Global Oscillation Network Group (GONG) Project}}.
\bjtitle{Science}
\bvolume{272},
\bfpage{1284}\,--\,\blpage{1286}.
doi:\doiurl{10.1126/science.272.5266.1284}.
\end{barticle}
\endbibitem

\bibitem[\protect\citeauthoryear{{Heinzel} and {Anzer}}{2001}]{hei01}
\begin{barticle}
\bauthor{\bsnm{{Heinzel}}, \binits{P.}}, \bauthor{\bsnm{{Anzer}}, \binits{U.}}:
\byear{2001},
\batitle{{Prominence fine structures in a magnetic equilibrium: Two-dimensional
  models with multilevel radiative transfer}}.
\bjtitle{\aap}
\bvolume{375},
\bfpage{1082}\,--\,\blpage{1090}.
doi:\doiurl{10.1051/0004-6361:20010926}.
\end{barticle}
\endbibitem

\bibitem[\protect\citeauthoryear{{Heinzel}, {Anzer}, and
  {Schmieder}}{2003}]{hei03}
\begin{barticle}
\bauthor{\bsnm{{Heinzel}}, \binits{P.}}, \bauthor{\bsnm{{Anzer}}, \binits{U.}},
  \bauthor{\bsnm{{Schmieder}}, \binits{B.}}:
\byear{2003},
\batitle{{A Spectroscopic Model of euv Filaments}}.
\bjtitle{\solphys}
\bvolume{216},
\bfpage{159}\,--\,\blpage{171}.
doi:\doiurl{10.1023/A:1026130028966}.
\end{barticle}
\endbibitem

\bibitem[\protect\citeauthoryear{{Heinzel} \textit{et~al.}}{2008}]{hei08}
\begin{barticle}
\bauthor{\bsnm{{Heinzel}}, \binits{P.}}, \bauthor{\bsnm{{Schmieder}},
  \binits{B.}}, \bauthor{\bsnm{{F{\'a}rn{\'{\i}}k}}, \binits{F.}},
  \bauthor{\bsnm{{Schwartz}}, \binits{P.}}, \bauthor{\bsnm{{Labrosse}},
  \binits{N.}}, \bauthor{\bsnm{{Kotr{\v c}}}, \binits{P.}},
  \bauthor{\bsnm{{Anzer}}, \binits{U.}}, \bauthor{\bsnm{{Molodij}},
  \binits{G.}}, \bauthor{\bsnm{{Berlicki}}, \binits{A.}},
  \bauthor{\bsnm{{DeLuca}}, \binits{E.E.}}, \bauthor{\bsnm{{Golub}},
  \binits{L.}}, \bauthor{\bsnm{{Watanabe}}, \binits{T.}},
  \bauthor{\bsnm{{Berger}}, \binits{T.}}:
\byear{2008},
\batitle{{Hinode, TRACE, SOHO, and Ground-based Observations of a Quiescent
  Prominence}}.
\bjtitle{\apj}
\bvolume{686},
\bfpage{1383}\,--\,\blpage{1396}.
doi:\doiurl{10.1086/591018}.
\end{barticle}
\endbibitem

\bibitem[\protect\citeauthoryear{{Hirayama}}{1985}]{hir85}
\begin{barticle}
\bauthor{\bsnm{{Hirayama}}, \binits{T.}}:
\byear{1985},
\batitle{{Modern observations of solar prominences}}.
\bjtitle{\solphys}
\bvolume{100},
\bfpage{415}\,--\,\blpage{434}.
doi:\doiurl{10.1007/BF00158439}.
\end{barticle}
\endbibitem

\bibitem[\protect\citeauthoryear{{Howard} \textit{et~al.}}{2008}]{Howard08}
\begin{barticle}
\bauthor{\bsnm{{Howard}}, \binits{R.A.}}, \bauthor{\bsnm{{Moses}},
  \binits{J.D.}}, \bauthor{\bsnm{{Vourlidas}}, \binits{A.}},
  \bauthor{\bsnm{{Newmark}}, \binits{J.S.}}, \bauthor{\bsnm{{Socker}},
  \binits{D.G.}}, \bauthor{\bsnm{{Plunkett}}, \binits{S.P.}},
  \bauthor{\bsnm{{Korendyke}}, \binits{C.M.}}, \bauthor{\bsnm{{Cook}},
  \binits{J.W.}}, \bauthor{\bsnm{{Hurley}}, \binits{A.}},
  \bauthor{\bsnm{{Davila}}, \binits{J.M.}}, \bauthor{\bsnm{{Thompson}},
  \binits{W.T.}}, \bauthor{\bsnm{{St Cyr}}, \binits{O.C.}},
  \bauthor{\bsnm{{Mentzell}}, \binits{E.}}, \bauthor{\bsnm{{Mehalick}},
  \binits{K.}}, \bauthor{\bsnm{{Lemen}}, \binits{J.R.}},
  \bauthor{\bsnm{{Wuelser}}, \binits{J.P.}}, \bauthor{\bsnm{{Duncan}},
  \binits{D.W.}}, \bauthor{\bsnm{{Tarbell}}, \binits{T.D.}},
  \bauthor{\bsnm{{Wolfson}}, \binits{C.J.}}, \bauthor{\bsnm{{Moore}},
  \binits{A.}}, \bauthor{\bsnm{{Harrison}}, \binits{R.A.}},
  \bauthor{\bsnm{{Waltham}}, \binits{N.R.}}, \bauthor{\bsnm{{Lang}},
  \binits{J.}}, \bauthor{\bsnm{{Davis}}, \binits{C.J.}},
  \bauthor{\bsnm{{Eyles}}, \binits{C.J.}}, \bauthor{\bsnm{{Mapson-Menard}},
  \binits{H.}}, \bauthor{\bsnm{{Simnett}}, \binits{G.M.}},
  \bauthor{\bsnm{{Halain}}, \binits{J.P.}}, \bauthor{\bsnm{{Defise}},
  \binits{J.M.}}, \bauthor{\bsnm{{Mazy}}, \binits{E.}},
  \bauthor{\bsnm{{Rochus}}, \binits{P.}}, \bauthor{\bsnm{{Mercier}},
  \binits{R.}}, \bauthor{\bsnm{{Ravet}}, \binits{M.F.}},
  \bauthor{\bsnm{{Delmotte}}, \binits{F.}}, \bauthor{\bsnm{{Auchere}},
  \binits{F.}}, \bauthor{\bsnm{{Delaboudiniere}}, \binits{J.P.}},
  \bauthor{\bsnm{{Bothmer}}, \binits{V.}}, \bauthor{\bsnm{{Deutsch}},
  \binits{W.}}, \bauthor{\bsnm{{Wang}}, \binits{D.}}, \bauthor{\bsnm{{Rich}},
  \binits{N.}}, \bauthor{\bsnm{{Cooper}}, \binits{S.}},
  \bauthor{\bsnm{{Stephens}}, \binits{V.}}, \bauthor{\bsnm{{Maahs}},
  \binits{G.}}, \bauthor{\bsnm{{Baugh}}, \binits{R.}},
  \bauthor{\bsnm{{McMullin}}, \binits{D.}}, \bauthor{\bsnm{{Carter}},
  \binits{T.}}:
\byear{2008},
\batitle{{Sun Earth Connection Coronal and Heliospheric Investigation
  (SECCHI)}}.
\bjtitle{\ssr}
\bvolume{136},
\bfpage{67}\,--\,\blpage{115}.
doi:\doiurl{10.1007/s11214-008-9341-4}.
\end{barticle}
\endbibitem

\bibitem[\protect\citeauthoryear{{Hudson} \textit{et~al.}}{1999}]{Hudson99}
\begin{barticle}
\bauthor{\bsnm{{Hudson}}, \binits{H.S.}}, \bauthor{\bsnm{{Acton}},
  \binits{L.W.}}, \bauthor{\bsnm{{Harvey}}, \binits{K.L.}},
  \bauthor{\bsnm{{McKenzie}}, \binits{D.E.}}:
\byear{1999},
\batitle{{A Stable Filament Cavity with a Hot Core}}.
\bjtitle{\apj}
\bvolume{513},
\bfpage{L83}\,--\,\blpage{L86}.
doi:\doiurl{10.1086/311892}.
\end{barticle}
\endbibitem

\bibitem[\protect\citeauthoryear{{Kucera}, {Andretta}, and
  {Poland}}{1998}]{kuc98}
\begin{barticle}
\bauthor{\bsnm{{Kucera}}, \binits{T.A.}}, \bauthor{\bsnm{{Andretta}},
  \binits{V.}}, \bauthor{\bsnm{{Poland}}, \binits{A.I.}}:
\byear{1998},
\batitle{{Neutral Hydrogen Column Depths in Prominences Using EUV Absorption
  Features}}.
\bjtitle{\solphys}
\bvolume{183},
\bfpage{107}\,--\,\blpage{121}.
\end{barticle}
\endbibitem

\bibitem[\protect\citeauthoryear{{Kucera}, {Tovar}, and {de
  Pontieu}}{2003}]{kuc03}
\begin{barticle}
\bauthor{\bsnm{{Kucera}}, \binits{T.A.}}, \bauthor{\bsnm{{Tovar}},
  \binits{M.}}, \bauthor{\bsnm{{de Pontieu}}, \binits{B.}}:
\byear{2003},
\batitle{{Prominence Motions Observed at High Cadences in Temperatures from 10
  000 to 250 000 K}}.
\bjtitle{\solphys}
\bvolume{212},
\bfpage{81}\,--\,\blpage{97}.
doi:\doiurl{10.1023/A:1022900604972}.
\end{barticle}
\endbibitem

\bibitem[\protect\citeauthoryear{{Kuperus}}{1996}]{kup96}
\begin{barticle}
\bauthor{\bsnm{{Kuperus}}, \binits{M.}}:
\byear{1996},
\batitle{{The Double Inverse Polarity Paradigm|The Sign of Magnetic Fields in
  Quiescent Prominences}}.
\bjtitle{\solphys}
\bvolume{169},
\bfpage{349}\,--\,\blpage{356}.
doi:\doiurl{10.1007/BF00190611}.
\end{barticle}
\endbibitem

\bibitem[\protect\citeauthoryear{{Kuperus} and {Raadu}}{1974}]{kup74}
\begin{barticle}
\bauthor{\bsnm{{Kuperus}}, \binits{M.}}, \bauthor{\bsnm{{Raadu}},
  \binits{M.A.}}:
\byear{1974},
\batitle{{The Support of Prominences Formed in Neutral Sheets}}.
\bjtitle{\aap}
\bvolume{31},
\bfpage{189}\,--\,\blpage{193}.
\end{barticle}
\endbibitem

\bibitem[\protect\citeauthoryear{{Labrosse} \textit{et~al.}}{2010}]{lab10}
\begin{barticle}
\bauthor{\bsnm{{Labrosse}}, \binits{N.}}, \bauthor{\bsnm{{Heinzel}},
  \binits{P.}}, \bauthor{\bsnm{{Vial}}, \binits{J.C.}},
  \bauthor{\bsnm{{Kucera}}, \binits{T.}}, \bauthor{\bsnm{{Parenti}},
  \binits{S.}}, \bauthor{\bsnm{{Gun{\'a}r}}, \binits{S.}},
  \bauthor{\bsnm{{Schmieder}}, \binits{B.}}, \bauthor{\bsnm{{Kilper}},
  \binits{G.}}:
\byear{2010},
\batitle{{Physics of Solar Prominences: Spectral Diagnostics and Non-LTE
  Modelling}}.
\bjtitle{\ssr}
\bvolume{151},
\bfpage{243}\,--\,\blpage{332}.
doi:\doiurl{10.1007/s11214-010-9630-6}.
\end{barticle}
\endbibitem

\bibitem[\protect\citeauthoryear{{Lemen} \textit{et~al.}}{2012}]{lem12}
\begin{barticle}
\bauthor{\bsnm{{Lemen}}, \binits{J.R.}}, \bauthor{\bsnm{{Title}},
  \binits{A.M.}}, \bauthor{\bsnm{{Akin}}, \binits{D.J.}},
  \bauthor{\bsnm{{Boerner}}, \binits{P.F.}}, \bauthor{\bsnm{{Chou}},
  \binits{C.}}, \bauthor{\bsnm{{Drake}}, \binits{J.F.}},
  \bauthor{\bsnm{{Duncan}}, \binits{D.W.}}, \bauthor{\bsnm{{Edwards}},
  \binits{C.G.}}, \bauthor{\bsnm{{Friedlaender}}, \binits{F.M.}},
  \bauthor{\bsnm{{Heyman}}, \binits{G.F.}}, \bauthor{\bsnm{{Hurlburt}},
  \binits{N.E.}}, \bauthor{\bsnm{{Katz}}, \binits{N.L.}},
  \bauthor{\bsnm{{Kushner}}, \binits{G.D.}}, \bauthor{\bsnm{{Levay}},
  \binits{M.}}, \bauthor{\bsnm{{Lindgren}}, \binits{R.W.}},
  \bauthor{\bsnm{{Mathur}}, \binits{D.P.}}, \bauthor{\bsnm{{McFeaters}},
  \binits{E.L.}}, \bauthor{\bsnm{{Mitchell}}, \binits{S.}},
  \bauthor{\bsnm{{Rehse}}, \binits{R.A.}}, \bauthor{\bsnm{{Schrijver}},
  \binits{C.J.}}, \bauthor{\bsnm{{Springer}}, \binits{L.A.}},
  \bauthor{\bsnm{{Stern}}, \binits{R.A.}}, \bauthor{\bsnm{{Tarbell}},
  \binits{T.D.}}, \bauthor{\bsnm{{Wuelser}}, \binits{J.P.}},
  \bauthor{\bsnm{{Wolfson}}, \binits{C.J.}}, \bauthor{\bsnm{{Yanari}},
  \binits{C.}}, \bauthor{\bsnm{{Bookbinder}}, \binits{J.A.}},
  \bauthor{\bsnm{{Cheimets}}, \binits{P.N.}}, \bauthor{\bsnm{{Caldwell}},
  \binits{D.}}, \bauthor{\bsnm{{Deluca}}, \binits{E.E.}},
  \bauthor{\bsnm{{Gates}}, \binits{R.}}, \bauthor{\bsnm{{Golub}}, \binits{L.}},
  \bauthor{\bsnm{{Park}}, \binits{S.}}, \bauthor{\bsnm{{Podgorski}},
  \binits{W.A.}}, \bauthor{\bsnm{{Bush}}, \binits{R.I.}},
  \bauthor{\bsnm{{Scherrer}}, \binits{P.H.}}, \bauthor{\bsnm{{Gummin}},
  \binits{M.A.}}, \bauthor{\bsnm{{Smith}}, \binits{P.}},
  \bauthor{\bsnm{{Auker}}, \binits{G.}}, \bauthor{\bsnm{{Jerram}},
  \binits{P.}}, \bauthor{\bsnm{{Pool}}, \binits{P.}}, \bauthor{\bsnm{{Soufli}},
  \binits{R.}}, \bauthor{\bsnm{{Windt}}, \binits{D.L.}},
  \bauthor{\bsnm{{Beardsley}}, \binits{S.}}, \bauthor{\bsnm{{Clapp}},
  \binits{M.}}, \bauthor{\bsnm{{Lang}}, \binits{J.}},
  \bauthor{\bsnm{{Waltham}}, \binits{N.}}:
\byear{2012},
\batitle{{The Atmospheric Imaging Assembly (AIA) on the Solar Dynamics
  Observatory (SDO)}}.
\bjtitle{\solphys}
\bvolume{275},
\bfpage{17}\,--\,\blpage{40}.
doi:\doiurl{10.1007/s11207-011-9776-8}.
\end{barticle}
\endbibitem

\bibitem[\protect\citeauthoryear{{Leroy}, {Bommier}, and
  {Sahal-Brechot}}{1984}]{leroy84}
\begin{barticle}
\bauthor{\bsnm{{Leroy}}, \binits{J.L.}}, \bauthor{\bsnm{{Bommier}},
  \binits{V.}}, \bauthor{\bsnm{{Sahal-Brechot}}, \binits{S.}}:
\byear{1984},
\batitle{{New data on the magnetic structure of quiescent prominences}}.
\bjtitle{\aap}
\bvolume{131},
\bfpage{33}\,--\,\blpage{44}.
\end{barticle}
\endbibitem

\bibitem[\protect\citeauthoryear{{Li} and {Zhang}}{2013}]{li2013}
\begin{barticle}
\bauthor{\bsnm{{Li}}, \binits{L.}}, \bauthor{\bsnm{{Zhang}}, \binits{J.}}:
\byear{2013},
\batitle{{The Evolution of Barbs of a Polar Crown Filament Observed by SDO}}.
\bjtitle{\solphys}
\bvolume{282},
\bfpage{147}\,--\,\blpage{174}.
doi:\doiurl{10.1007/s11207-012-0122-6}.
\end{barticle}
\endbibitem

\bibitem[\protect\citeauthoryear{{Lin}, {Engvold}, and {Wiik}}{2003}]{lin2003}
\begin{barticle}
\bauthor{\bsnm{{Lin}}, \binits{Y.}}, \bauthor{\bsnm{{Engvold}}, \binits{O.R.}},
  \bauthor{\bsnm{{Wiik}}, \binits{J.E.}}:
\byear{2003},
\batitle{{Counterstreaming in a Large Polar Crown Filament}}.
\bjtitle{\solphys}
\bvolume{216},
\bfpage{109}\,--\,\blpage{120}.
doi:\doiurl{10.1023/A:1026150809598}.
\end{barticle}
\endbibitem

\bibitem[\protect\citeauthoryear{{Low} and {Hundhausen}}{1995}]{low95}
\begin{barticle}
\bauthor{\bsnm{{Low}}, \binits{B.C.}}, \bauthor{\bsnm{{Hundhausen}},
  \binits{J.R.}}:
\byear{1995},
\batitle{{Magnetostatic structures of the solar corona. 2: The magnetic
  topology of quiescent prominences}}.
\bjtitle{\apj}
\bvolume{443},
\bfpage{818}\,--\,\blpage{836}.
doi:\doiurl{10.1086/175572}.
\end{barticle}
\endbibitem

\bibitem[\protect\citeauthoryear{{Low} and {Zhang}}{2002}]{low02}
\begin{barticle}
\bauthor{\bsnm{{Low}}, \binits{B.C.}}, \bauthor{\bsnm{{Zhang}}, \binits{M.}}:
\byear{2002},
\batitle{{The Hydromagnetic Origin of the Two Dynamical Types of Solar Coronal
  Mass Ejections}}.
\bjtitle{\apjl}
\bvolume{564},
\bfpage{L53}\,--\,\blpage{L56}.
doi:\doiurl{10.1086/338798}.
\end{barticle}
\endbibitem

\bibitem[\protect\citeauthoryear{{Mackay} and {van Ballegooijen}}{2001}]{mac01}
\begin{barticle}
\bauthor{\bsnm{{Mackay}}, \binits{D.H.}}, \bauthor{\bsnm{{van Ballegooijen}},
  \binits{A.A.}}:
\byear{2001},
\batitle{{A Possible Solar Cycle Dependence to the Hemispheric Pattern of
  Filament Magnetic Fields?}}
\bjtitle{\apj}
\bvolume{560},
\bfpage{445}\,--\,\blpage{455}.
doi:\doiurl{10.1086/322385}.
\end{barticle}
\endbibitem

\bibitem[\protect\citeauthoryear{{Mackay} and {van Ballegooijen}}{2006}]{mac06}
\begin{barticle}
\bauthor{\bsnm{{Mackay}}, \binits{D.H.}}, \bauthor{\bsnm{{van Ballegooijen}},
  \binits{A.A.}}:
\byear{2006},
\batitle{{Models of the Large-Scale Corona. II. Magnetic Connectivity and Open
  Flux Variation}}.
\bjtitle{\apj}
\bvolume{642},
\bfpage{1193}\,--\,\blpage{1204}.
doi:\doiurl{10.1086/501043}.
\end{barticle}
\endbibitem

\bibitem[\protect\citeauthoryear{{Mackay}, {Gaizauskas}, and
  {Yeates}}{2008}]{Mackay08}
\begin{barticle}
\bauthor{\bsnm{{Mackay}}, \binits{D.H.}}, \bauthor{\bsnm{{Gaizauskas}},
  \binits{V.}}, \bauthor{\bsnm{{Yeates}}, \binits{A.R.}}:
\byear{2008},
\batitle{{Where Do Solar Filaments Form?: Consequences for Theoretical
  Models}}.
\bjtitle{\solphys}
\bvolume{248},
\bfpage{51}\,--\,\blpage{65}.
doi:\doiurl{10.1007/s11207-008-9127-6}.
\end{barticle}
\endbibitem

\bibitem[\protect\citeauthoryear{{Mackay} \textit{et~al.}}{2010}]{Mackay10}
\begin{barticle}
\bauthor{\bsnm{{Mackay}}, \binits{D.H.}}, \bauthor{\bsnm{{Karpen}},
  \binits{J.T.}}, \bauthor{\bsnm{{Ballester}}, \binits{J.L.}},
  \bauthor{\bsnm{{Schmieder}}, \binits{B.}}, \bauthor{\bsnm{{Aulanier}},
  \binits{G.}}:
\byear{2010},
\batitle{{Physics of Solar Prominences: II Magnetic Structure and Dynamics}}.
\bjtitle{\ssr}
\bvolume{151},
\bfpage{333}\,--\,\blpage{399}.
doi:\doiurl{10.1007/s11214-010-9628-0}.
\end{barticle}
\endbibitem

\bibitem[\protect\citeauthoryear{{Martens} and {Zwaan}}{2001}]{Martens01}
\begin{barticle}
\bauthor{\bsnm{{Martens}}, \binits{P.C.}}, \bauthor{\bsnm{{Zwaan}},
  \binits{C.}}:
\byear{2001},
\batitle{{Origin and Evolution of Filament-Prominence Systems}}.
\bjtitle{\apj}
\bvolume{558},
\bfpage{872}\,--\,\blpage{887}.
doi:\doiurl{10.1086/322279}.
\end{barticle}
\endbibitem

\bibitem[\protect\citeauthoryear{{Martin}}{1973}]{Martin73}
\begin{barticle}
\bauthor{\bsnm{{Martin}}, \binits{S.F.}}:
\byear{1973},
\batitle{{The Evolution of Prominences and Their Relationship to Active Centers
  (A Review)}}.
\bjtitle{\solphys}
\bvolume{31},
\bfpage{3}\,--\,\blpage{21}.
doi:\doiurl{10.1007/BF00156070}.
\end{barticle}
\endbibitem

\bibitem[\protect\citeauthoryear{{Martin}}{1998}]{Martin98}
\begin{barticle}
\bauthor{\bsnm{{Martin}}, \binits{S.F.}}:
\byear{1998},
\batitle{{Conditions for the Formation and Maintenance of Filaments (Invited
  Review)}}.
\bjtitle{\solphys}
\bvolume{182},
\bfpage{107}\,--\,\blpage{137}.
doi:\doiurl{10.1023/A:1005026814076}.
\end{barticle}
\endbibitem

\bibitem[\protect\citeauthoryear{{McIntosh}}{1992}]{McIntosh92}
\begin{botherref}
\oauthor{\bsnm{{McIntosh}}, \binits{P.S.}}:
1992,
{Solar Interior Processes Suggested by Large-Scale Surface Patterns}.
In: {K.~L.~Harvey} (ed.)
\textit{The Solar Cycle},
\textit{ASP}
\textbf{\textbf{CS-27}},
14\,--\,34.
\end{botherref}
\endbibitem

\bibitem[\protect\citeauthoryear{{Minarovjech}, {Rybansky}, and
  {Rusin}}{1998}]{Minar98}
\begin{botherref}
\oauthor{\bsnm{{Minarovjech}}, \binits{M.}}, \oauthor{\bsnm{{Rybansky}},
  \binits{M.}}, \oauthor{\bsnm{{Rusin}}, \binits{V.}}:
1998,
{Prominences and the Green Corona Over the Solar Activity Cycle}
\textbf{177},
357\,--\,364.
\end{botherref}
\endbibitem

\bibitem[\protect\citeauthoryear{{Panesar} \textit{et~al.}}{2013}]{panesar13}
\begin{barticle}
\bauthor{\bsnm{{Panesar}}, \binits{N.K.}}, \bauthor{\bsnm{{Innes}},
  \binits{D.E.}}, \bauthor{\bsnm{{Tiwari}}, \binits{S.K.}},
  \bauthor{\bsnm{{Low}}, \binits{B.C.}}:
\byear{2013},
\batitle{{A solar tornado triggered by flares?}}
\bjtitle{\aap}
\bvolume{549},
\bfpage{A105}.
doi:\doiurl{10.1051/0004-6361/201220503}.
\end{barticle}
\endbibitem

\bibitem[\protect\citeauthoryear{{Parenti} \textit{et~al.}}{2012}]{par12}
\begin{barticle}
\bauthor{\bsnm{{Parenti}}, \binits{S.}}, \bauthor{\bsnm{{Schmieder}},
  \binits{B.}}, \bauthor{\bsnm{{Heinzel}}, \binits{P.}},
  \bauthor{\bsnm{{Golub}}, \binits{L.}}:
\byear{2012},
\batitle{{On the Nature of Prominence Emission Observed by SDO/AIA}}.
\bjtitle{\apj}
\bvolume{754},
\bfpage{66}.
doi:\doiurl{10.1088/0004-637X/754/1/66}.
\end{barticle}
\endbibitem

\bibitem[\protect\citeauthoryear{{Priest}, {Hood}, and
  {Anzer}}{1989}]{Priest89}
\begin{barticle}
\bauthor{\bsnm{{Priest}}, \binits{E.R.}}, \bauthor{\bsnm{{Hood}},
  \binits{A.W.}}, \bauthor{\bsnm{{Anzer}}, \binits{U.}}:
\byear{1989},
\batitle{{A twisted flux-tube model for solar prominences. I - General
  properties}}.
\bjtitle{\apj}
\bvolume{344},
\bfpage{1010}\,--\,\blpage{1025}.
doi:\doiurl{10.1086/167868}.
\end{barticle}
\endbibitem

\bibitem[\protect\citeauthoryear{{Rachmeler} \textit{et~al.}}{2013}]{rach13}
\begin{barticle}
\bauthor{\bsnm{{Rachmeler}}, \binits{L.A.}}, \bauthor{\bsnm{{Gibson}},
  \binits{S.E.}}, \bauthor{\bsnm{{Dove}}, \binits{J.B.}},
  \bauthor{\bsnm{{DeVore}}, \binits{C.R.}}, \bauthor{\bsnm{{Fan}},
  \binits{Y.}}:
\byear{2013},
\batitle{{Polarimetric Properties of Flux Ropes and Sheared Arcades in Coronal
  Prominence Cavities}}.
\bjtitle{\solphys}
\bvolume{288},
\bfpage{617}\,--\,\blpage{636}.
\end{barticle}
\endbibitem

\bibitem[\protect\citeauthoryear{{R{\'e}gnier}, {Walsh}, and
  {Alexander}}{2011}]{Regnier11}
\begin{barticle}
\bauthor{\bsnm{{R{\'e}gnier}}, \binits{S.}}, \bauthor{\bsnm{{Walsh}},
  \binits{R.W.}}, \bauthor{\bsnm{{Alexander}}, \binits{C.E.}}:
\byear{2011},
\batitle{{A new look at a polar crown cavity as observed by SDO/AIA. Structure
  and dynamics}}.
\bjtitle{\aa}
\bvolume{533},
\bfpage{L1}.
doi:\doiurl{10.1051/0004-6361/201117381}.
\end{barticle}
\endbibitem

\bibitem[\protect\citeauthoryear{{Schmieder} \textit{et~al.}}{2004}]{sch04}
\begin{barticle}
\bauthor{\bsnm{{Schmieder}}, \binits{B.}}, \bauthor{\bsnm{{Lin}}, \binits{Y.}},
  \bauthor{\bsnm{{Heinzel}}, \binits{P.}}, \bauthor{\bsnm{{Schwartz}},
  \binits{P.}}:
\byear{2004},
\batitle{{Multi-wavelength study of a high-latitude EUV filament}}.
\bjtitle{\solphys}
\bvolume{221},
\bfpage{297}\,--\,\blpage{323}.
doi:\doiurl{10.1023/B:SOLA.0000035059.50427.68}.
\end{barticle}
\endbibitem

\bibitem[\protect\citeauthoryear{{Schmieder} \textit{et~al.}}{2010}]{sch10}
\begin{barticle}
\bauthor{\bsnm{{Schmieder}}, \binits{B.}}, \bauthor{\bsnm{{Chandra}},
  \binits{R.}}, \bauthor{\bsnm{{Berlicki}}, \binits{A.}},
  \bauthor{\bsnm{{Mein}}, \binits{P.}}:
\byear{2010},
\batitle{{Velocity vectors of a quiescent prominence observed by Hinode/SOT and
  the MSDP (Meudon)}}.
\bjtitle{\aap}
\bvolume{514},
\bfpage{A68}.
doi:\doiurl{10.1051/0004-6361/200913477}.
\end{barticle}
\endbibitem

\bibitem[\protect\citeauthoryear{{Schmit} and {Gibson}}{2013}]{schmit13}
\begin{barticle}
\bauthor{\bsnm{{Schmit}}, \binits{D.J.}}, \bauthor{\bsnm{{Gibson}},
  \binits{S.}}:
\byear{2013},
\batitle{{Diagnosing the Prominence-Cavity Connection}}.
\bjtitle{\apj}
\bvolume{770},
\bfpage{35}.
doi:\doiurl{10.1088/0004-637X/770/1/35}.
\end{barticle}
\endbibitem

\bibitem[\protect\citeauthoryear{{Schmit} and {Gibson}}{2011}]{don11}
\begin{barticle}
\bauthor{\bsnm{{Schmit}}, \binits{D.J.}}, \bauthor{\bsnm{{Gibson}},
  \binits{S.E.}}:
\byear{2011},
\batitle{{Forward Modeling Cavity Density: A Multi-instrument Diagnostic}}.
\bjtitle{\apj}
\bvolume{733},
\bfpage{1}.
doi:\doiurl{10.1088/0004-637X/733/1/1}.
\end{barticle}
\endbibitem

\bibitem[\protect\citeauthoryear{{Schou} \textit{et~al.}}{2012}]{sch12}
\begin{barticle}
\bauthor{\bsnm{{Schou}}, \binits{J.}}, \bauthor{\bsnm{{Scherrer}},
  \binits{P.H.}}, \bauthor{\bsnm{{Bush}}, \binits{R.I.}},
  \bauthor{\bsnm{{Wachter}}, \binits{R.}}, \bauthor{\bsnm{{Couvidat}},
  \binits{S.}}, \bauthor{\bsnm{{Rabello-Soares}}, \binits{M.C.}},
  \bauthor{\bsnm{{Bogart}}, \binits{R.S.}}, \bauthor{\bsnm{{Hoeksema}},
  \binits{J.T.}}, \bauthor{\bsnm{{Liu}}, \binits{Y.}},
  \bauthor{\bsnm{{Duvall}}, \binits{T.L.}}, \bauthor{\bsnm{{Akin}},
  \binits{D.J.}}, \bauthor{\bsnm{{Allard}}, \binits{B.A.}},
  \bauthor{\bsnm{{Miles}}, \binits{J.W.}}, \bauthor{\bsnm{{Rairden}},
  \binits{R.}}, \bauthor{\bsnm{{Shine}}, \binits{R.A.}},
  \bauthor{\bsnm{{Tarbell}}, \binits{T.D.}}, \bauthor{\bsnm{{Title}},
  \binits{A.M.}}, \bauthor{\bsnm{{Wolfson}}, \binits{C.J.}},
  \bauthor{\bsnm{{Elmore}}, \binits{D.F.}}, \bauthor{\bsnm{{Norton}},
  \binits{A.A.}}, \bauthor{\bsnm{{Tomczyk}}, \binits{S.}}:
\byear{2012},
\batitle{{Design and Ground Calibration of the Helioseismic and Magnetic Imager
  (HMI) Instrument on the Solar Dynamics Observatory (SDO)}}.
\bjtitle{\solphys}
\bvolume{275},
\bfpage{229}\,--\,\blpage{259}.
doi:\doiurl{10.1007/s11207-011-9842-2}.
\end{barticle}
\endbibitem

\bibitem[\protect\citeauthoryear{{Su} and {van Ballegooijen}}{2012}]{su12}
\begin{barticle}
\bauthor{\bsnm{{Su}}, \binits{Y.}}, \bauthor{\bsnm{{van Ballegooijen}},
  \binits{A.}}:
\byear{2012},
\batitle{{Observations and Magnetic Field Modeling of a Solar Polar Crown
  Prominence}}.
\bjtitle{\apj}
\bvolume{757},
\bfpage{168}.
doi:\doiurl{10.1088/0004-637X/757/2/168}.
\end{barticle}
\endbibitem

\bibitem[\protect\citeauthoryear{{Tandberg-Hanssen}}{1995}]{tand95}
\begin{botherref}
\oauthor{\bsnm{{Tandberg-Hanssen}}, \binits{E.}} (ed.):
1995,
\textit{{The nature of solar prominences}},
\textit{Astrophysics and Space Science Library}
\textbf{199}.
\end{botherref}
\endbibitem

\bibitem[\protect\citeauthoryear{{van Ballegooijen} and
  {Martens}}{1989}]{Balle89}
\begin{barticle}
\bauthor{\bsnm{{van Ballegooijen}}, \binits{A.A.}}, \bauthor{\bsnm{{Martens}},
  \binits{P.C.H.}}:
\byear{1989},
\batitle{{Formation and eruption of solar prominences}}.
\bjtitle{\apj}
\bvolume{343},
\bfpage{971}\,--\,\blpage{984}.
doi:\doiurl{10.1086/167766}.
\end{barticle}
\endbibitem

\bibitem[\protect\citeauthoryear{{V{\'a}squez}, {Frazin}, and
  {Kamalabadi}}{2009}]{vas09}
\begin{barticle}
\bauthor{\bsnm{{V{\'a}squez}}, \binits{A.M.}}, \bauthor{\bsnm{{Frazin}},
  \binits{R.A.}}, \bauthor{\bsnm{{Kamalabadi}}, \binits{F.}}:
\byear{2009},
\batitle{{3D Temperatures and Densities of the Solar Corona via
  Multi-Spacecraft EUV Tomography: Analysis of Prominence Cavities}}.
\bjtitle{\solphys}
\bvolume{256},
\bfpage{73}\,--\,\blpage{85}.
doi:\doiurl{10.1007/s11207-009-9321-1}.
\end{barticle}
\endbibitem

\bibitem[\protect\citeauthoryear{{Wang} \textit{et~al.}}{1998}]{wangh98}
\begin{barticle}
\bauthor{\bsnm{{Wang}}, \binits{H.}}, \bauthor{\bsnm{{Chae}}, \binits{J.}},
  \bauthor{\bsnm{{Gurman}}, \binits{J.B.}}, \bauthor{\bsnm{{Kucera}},
  \binits{T.A.}}:
\byear{1998},
\batitle{{Comparison of Prominences in H{$\alpha$} and He II 304 {\AA}}}.
\bjtitle{\solphys}
\bvolume{183},
\bfpage{91}\,--\,\blpage{96}.
doi:\doiurl{10.1023/A:1005010504873}.
\end{barticle}
\endbibitem

\bibitem[\protect\citeauthoryear{{Wang}}{1999}]{wan99}
\begin{barticle}
\bauthor{\bsnm{{Wang}}, \binits{Y.M.}}:
\byear{1999},
\batitle{{The Jetlike Nature of HE II $\lambda$ 304 Prominences}}.
\bjtitle{\apj}
\bvolume{520},
\bfpage{L71}\,--\,\blpage{L74}.
doi:\doiurl{10.1086/312149}.
\end{barticle}
\endbibitem

\bibitem[\protect\citeauthoryear{{Webb}, {Davis}, and
  {McIntosh}}{1984}]{Webb84}
\begin{barticle}
\bauthor{\bsnm{{Webb}}, \binits{D.F.}}, \bauthor{\bsnm{{Davis}},
  \binits{J.M.}}, \bauthor{\bsnm{{McIntosh}}, \binits{P.S.}}:
\byear{1984},
\batitle{{Observations of the reappearance of polar coronal holes and the
  reversal of the polar magnetic field}}.
\bjtitle{\solphys}
\bvolume{92},
\bfpage{109}\,--\,\blpage{132}.
doi:\doiurl{10.1007/BF00157239}.
\end{barticle}
\endbibitem

\bibitem[\protect\citeauthoryear{{Zirker} \textit{et~al.}}{1997}]{zir97}
\begin{barticle}
\bauthor{\bsnm{{Zirker}}, \binits{J.B.}}, \bauthor{\bsnm{{Martin}},
  \binits{S.F.}}, \bauthor{\bsnm{{Harvey}}, \binits{K.}},
  \bauthor{\bsnm{{Gaizauskas}}, \binits{V.}}:
\byear{1997},
\batitle{{Global Magnetic Patterns of Chirality}}.
\bjtitle{\solphys}
\bvolume{175},
\bfpage{27}\,--\,\blpage{44}.
\end{barticle}
\endbibitem

\end{thebibliography}

\end{article}

\end{document}